\documentclass[11pt,document,nofootinbib,superscriptaddress,onecolumn,preprintnumbers,balancelastpage]{article}
\pdfoutput=1
\usepackage{jheppub}

\usepackage{graphicx}
\usepackage{amsmath}
\usepackage{amsmath}
\usepackage{xspace}
\usepackage{subfig}
\usepackage{xcolor}
\usepackage{mathtools,slashed}
\usepackage{braket}
\usepackage{multirow}
\usepackage{slashed}

\usepackage{array}
\usepackage{multirow}
\newcommand{\s} {\scriptscriptstyle}

\title{
Probing Heavy Neutrino Magnetic Moments at the LHC using Long-Lived Particle Searches
}
\author[a]{Rebeca Beltr\'{a}n,}
\author[b,c]{Patrick D. Bolton,}
\author[d]{Frank F. Deppisch,}
\author[a]{Chandan Hati,}
\author[a]{Martin Hirsch}

\emailAdd{rebeca.beltran@ific.uv.es}
\emailAdd{patrick.bolton@ijs.si}
\emailAdd{f.deppisch@ucl.ac.uk}
\emailAdd{chandan@ific.uv.es}
\emailAdd{mahirsch@ific.uv.es}

\affiliation[a]{\footnotesize  Instituto de F\'{i}sica Corpuscular (IFIC), CSIC-Universitat de Valencia, C/ Catedrático José Beltrán 2, 46980, Valencia, Spain}
\affiliation[b]{\footnotesize SISSA, International School for Advanced Studies, INFN, Sezione di Trieste, Via Bonomea 265, I-34136 Trieste, Italy}
\affiliation[c]{Jožef Stefan Institute, Jamova 39, 1000 Ljubljana, Slovenia}
\affiliation[d]{\footnotesize  Department of Physics and Astronomy, University College London, London WC1E 6BT, United Kingdom}

\abstract{
We explore long-lived particle (LLP) searches using non-pointing photons at the LHC as a probe for sterile-to-sterile and active-to-sterile transition magnetic dipole moments of sterile neutrinos. We consider heavy sterile neutrinos with masses ranging from a few~GeV to several hundreds of GeV. We discuss transition magnetic dipole moments using the Standard Model effective field theory and low-energy effective field theory extended by sterile neutrinos ($N_R$SMEFT and $N_R$LEFT) and also provide a simplified UV-complete model example. LLP searches at the LHC using non-pointing photons will probe sterile-to-sterile dipole moments two orders of magnitude below the current best constraints from LEP, while an unprecedented sensitivity to sterile neutrino mass of about 700 GeV is expected for active-to-sterile dipole moments. For the UV model example with one-loop transition magnetic moments, the searches for charged lepton flavour violating processes in synergy with LLP searches at the LHC can probe new physics at several TeV mass scales and provide valuable insights into the lepton flavour structure of new physics couplings. 
}

\makeatletter
\gdef\@fpheader{\phantom{a}}
\makeatother

\begin{document}
\maketitle

\section{Introduction}
\label{sec:intro}

Neutrino oscillations currently provide the only concrete laboratory-based evidence of physics beyond the Standard Model~(SM). In the absence of direct signs of physics beyond the SM from resonance searches at the Large Hadron Collider~(LHC), it is therefore interesting to consider the possibility that the existence of new physics (NP) could manifest itself as non-trivial neutrino properties. One such intriguing attribute is the neutrino magnetic dipole moment. Minimally extending the SM to include non-zero active neutrino masses, the induced neutrino magnetic moments are tiny, $\mu_\nu \simeq 3 \times 10^{-20}\,\mu_B~(m_\nu/0.1~\text{eV})$, where $m_\nu$ is the neutrino mass scale and $\mu_B$ is the commonly employed unit of the Bohr magneton~\cite{Marciano:1977wx,Lee:1977tib,Fujikawa:1980yx}. Thus, a positive experimental signal of neutrino magnetic moments could provide a smoking gun for physics beyond the SM. 

Sizeable \textit{active-to-active} neutrino magnetic moments with naturally small neutrino masses have been a lively avenue of research in recent times~\cite{Shrock:1982sc,Voloshin:1987qy,Barbieri:1988fh,Babu:2020ivd,Babu:2021jnu}. With the addition of right-handed (RH) or sterile neutrinos to the SM field content, often invoked as a means to dynamically generate the light neutrino masses, the prospect of \textit{active-to-sterile} neutrino transition magnetic moments has also been of much interest in the literature~\cite{Magill:2018jla,Brdar:2020quo,Schwetz:2020xra,Bolton:2021pey,Fernandez-Martinez:2023phj}. A wide variety of flavour-dependent signatures of transition magnetic moments have been examined, such as the Primakoff upscattering of an active neutrino to a heavy state (with its subsequent decay to a light neutrino and a photon) in 
coherent elastic neutrino-nucleus scattering (CE$\nu$NS) experiments~\cite{Miranda:2019wdy, Li:2020lba, Miranda:2021kre, Bolton:2021pey}, NOMAD~\cite{Gninenko:1998nn,Magill:2018jla},
CHARM-II~\cite{Coloma:2017ppo}, 
DONUT~\cite{DONUT:2001zvi},
LSND~\cite{Magill:2018jla}, 
Borexino~\cite{Brdar:2020quo,Plestid:2020vqf}, 
XENON1T~\cite{Shoemaker:2018vii,Brdar:2020quo,Li:2022bqr},
IceCube~\cite{Huang:2022pce,Coloma:2017ppo},
MiniBooNE~\cite{Gninenko:2009ks,Magill:2018jla}, 
DUNE~\cite{Schwetz:2020xra,Atkinson:2021rnp,Ovchynnikov:2022rqj},
Super-Kamiokande~\cite{Plestid:2020vqf,Atkinson:2021rnp},
and FASER$\nu$~\cite{Jodlowski:2020vhr,Ismail:2021dyp}; the production of heavy states via meson decays, leading to invisible final states~\cite{Li:2020lba} and displaced vertex signatures~\cite{Barducci:2023hzo};
monophoton plus missing energy at LEP, LHC, and future colliders~\cite{Magill:2018jla, Zhang:2023nxy,Ovchynnikov:2023wgg,Barducci:2024kig}; 
modifications to Big Bang nucleosynthesis (BBN) affecting the abundance of $^{4}$He~\cite{Magill:2018jla}; 
and excess energy loss in SN1987A and low-energy supernovae~\cite{Magill:2018jla,Chauhan:2024nfa}. If active-to-sterile neutrino transition magnetic moments are discovered in the future, further observations may be used to discriminate the Dirac versus Majorana nature of the heavy states. For example, the photon emitted in the decay of a heavy state produced via upscattering in a CE$\nu$NS experiment exhibits different kinematic distributions in the two scenarios~\cite{Bolton:2021pey}. From the model-building perspective, large active-to-active and active-to-sterile neutrino transition magnetic moments are intimately connected with the radiative corrections to neutrino masses, requiring non-trivial 
symmetries or cancellations to realise a consistent ultraviolet (UV) completion~\cite{Bell:2005kz,Bell:2006wi}. 

On the other hand, \textit{sterile-to-sterile} neutrino transition magnetic moments remain relatively unconstrained for sterile neutrinos with masses above the heavy meson masses. The transition magnetic moments between GeV scale sterile neutrinos have been extensively studied in the context of long-lived particle (LLP) searches at AL3X, ANUBIS, CODEX-b, DUNE, FACET, FASER/FASER$\nu$, MAPP, MATHUSLA, and SHiP~\cite{Barducci:2022gdv, 
Barducci:2023hzo, Gunther:2023vmz}. For sterile neutrinos heavier than the energies accessible in meson decays, LLP searches at the LHC can provide an alternative window to sterile-to-sterile neutrino magnetic moments, as recently explored in~\cite{Liu:2023nxi,Chun:2024mus,Duarte:2023tdw}. In the context of $\text{CP}$-violation, the sterile-to-sterile neutrino magnetic moment has been discussed in~\cite{Balaji:2020oig}. The leading constraint in this mass regime (up to the $Z$ mass) comes from the monophoton plus missing energy signature at LEP \cite{Magill:2018jla}.

In this work, we start with the model-independent effective field theory (EFT) approach, examining the operators relevant for GeV to TeV scale neutrino magnetic moments within the frameworks of $N_R$SMEFT (SM effective field theory extended with RH neutrinos) and its low-energy counterpart, $N_R$LEFT (low-energy effective field theory extended with RH neutrinos). To complement this, we consider a UV-complete model example that generates these operators at the one-loop level. 
Our model is similar to the one proposed in~\cite{Aparici:2009oua}, but we generalise it further by removing the $Z_2$ symmetry imposed in the original construction. The broadening of the model parameter space leads to the interesting scenario where large active-to-sterile magnetic moments are also generated at one-loop, without relying on the highly suppressed active-sterile neutrino mixing. This opens up a wider range of phenomenology for LLP searches at the LHC. We also consider constraints from other probes on the UV model, such as electroweak precision observables and charged lepton flavour violation (cLFV).

We focus on the heavy sterile neutrino mass ranging from the GeV to several $100$~GeV to explore the constraints on the transition magnetic moments from LLP searches at the LHC using non-pointing photons. Our results for the simplified model example can be straightforwardly generalised to many extensions of the SM involving sterile neutrinos, an extended scalar sector, and vector-like fermions~\cite{Singer:1980sw,Babu:1989rb,Pilaftsis:1991ug,Deppisch:2017vne,Hati:2018tge,Hati:2018fzc,Deppisch:2016jzl,Fonseca:2016tbn,Hati:2020fzp}. We find that to probe the heavy neutrino dipole moments with a non-pointing photon final state (without any prompt charged particles) at the LHC, the transverse impact parameter can be used to select events with a displaced vertex. 
Our numerical simulations show that, depending on the dominant production and decay mechanisms for the sterile neutrinos, searches for LLPs with non-pointing photons will be sensitive to unexplored parts of the parameter space in sterile-to-sterile and active-to-sterile neutrino transition dipole moments for sterile neutrino mass ranging from tens of GeV to well beyond the $Z$ mass. 

The plan for the rest of the paper is as follows. In Section~\ref{sec:EFT}, we present an EFT description of the operators relevant to the sterile neutrino magnetic moment in the context of different EFTs. We then present a simple UV-complete model example for realising the sterile-to-sterile and active-to-sterile transition magnetic dipole moments in Section~\ref{sec:model}. In this Section, we also present the derivation of these magnetic moments and their correlation with neutrino masses. In Section~\ref{sec:constraints}, we review the various phenomenological constraints on the model from direct collider searches and searches for cLFV processes. In Section~\ref{sec:LLP}, we present different interesting final state signatures for probing the sterile-to-sterile and active-to-sterile transition magnetic dipole moments using LLP searches with non-pointing photons. We further discuss the implementation and simulation details. In Section~\ref{sec:results}, we present the numerical results in terms of the EFT coefficients and the parameters of the explicit UV model. Finally, we conclude in Section~\ref{sec:conclusions}. 

\section{Sterile Neutrino Transition Magnetic Moments}
\label{sec:EFT}

In the presence of RH or sterile neutrinos $N_R$ with masses below the electroweak (EW) scale, transition magnetic moments are described by the effective Lagrangian
\begin{align}
\label{Lag:LEFT$+N_R$}
\mathcal{L}_{N_R\text{LEFT}} &\supset d_{\s NN\gamma}^{\prime ij}\, \mathcal{O}_{\s NN\gamma}^{ij} + d_{\s \nu N\gamma}^{\prime \alpha i}\, \mathcal{O}_{\s \nu N\gamma}^{\alpha i} + \text{h.c.}\,,
\end{align}
with the dimension-five ($d = 5$) operators\footnote{While we omit it here, it is common to include a factor of $1/2$ in the definition of $\mathcal{O}_{\s NN\gamma}$ so that the normalisation is analogous to the Majorana and Dirac mass terms, as seen in Section~\ref{sec:neutrino_masses}.}
\begin{align}
\label{OP:LEFT$+N_R$}
\mathcal{O}_{\s NN\gamma}^{ij} = (\bar{N}_{Ri}^c\sigma_{\mu\nu}N_{Rj})F^{\mu\nu}\,,\quad
\mathcal{O}_{\s \nu N\gamma}^{\alpha i} =(\bar{\nu}_{L\alpha}\sigma_{\mu\nu}N_{Ri})F^{\mu\nu}\,,
\end{align}
where $\sigma_{\mu\nu} = \frac{i}{2}[\gamma_\mu,\gamma_\nu]$, $F_{\mu\nu}$ is the photon field strength and $N_R^c = \mathcal{C}\bar{N}_R^T$, with $\mathcal{C}$ the charge conjugation matrix. Here, we use $\alpha \in \{1,2,3\}$ and $i,j \in \{1,2,\ldots\}$ to denote the generation of SM lepton and sterile neutrino, respectively. The operators $\mathcal{O}_{\s NN\gamma}^{ij}$ and $\mathcal{O}_{\s \nu N\gamma}^{\alpha i}$ correspond to sterile-to-sterile and active-to-sterile neutrino transition magnetic moments, respectively. The former is antisymmetric in the indices $i,j$ and vanishes for a single sterile state. We note that these effective interactions are only applicable well below the EW symmetry breaking scale and are often referred to as $N_R$LEFT operators. We stress that the prime superscript is used to distinguish the $N_R$LEFT Wilson coefficients from the basis of operators obtained after rotating $N_R$SMEFT operators at the EW symmetry breaking scale, but not integrating out degrees of freedom such as $W$ and $Z$, which will be defined shortly.

The $N_R$LEFT operators in Eq.~\eqref{OP:LEFT$+N_R$} arise from operators above the EW scale in the $N_R$SMEFT, in particular, the operators (up to $d = 6$)
\begin{align}
\label{Lag:SMEFT$+N_R$}
\mathcal{L}_{N_R\text{SMEFT}} &\supset 
C^{(5)ij}_{\s NNB}\, \mathcal{O}_{\s NNB}^{(5)ij} +
C_{\s NB}^{(6)\alpha i} \, \mathcal{O}_{\s NB}^{(6)\alpha i}+ {C}_{\s NW}^{(6)\alpha i}\, \mathcal{O}_{\s NW}^{(6)\alpha i} + \text{h.c.} \,,
\end{align}
with
\begin{align}
\label{OP:SMEFT$+N_R$}
\mathcal{O}_{\s NNB}^{(5)ij} = (\bar{N}_{Ri}^c\sigma_{\mu\nu}N_{Rj})B^{\mu\nu} \,&, \\
\mathcal{O}_{\s NB}^{(6)\alpha i} = (\bar{L}_\alpha\sigma_{\mu\nu}N_{Ri})\tilde{H}B^{\mu\nu} \,&, \quad \mathcal{O}_{\s NW}^{(6)\alpha i} = (\bar{L}_\alpha\sigma_{\mu\nu}N_{Ri})\tau^I\tilde{H}W^{I\mu\nu} \,, 
\label{OP:SMEFT$+N_R$-dim6}
\end{align}
where $L = (\nu_L~\ell_L)^T$ and $H = (\varphi^+, \varphi^0)^T$ are the SM lepton and Higgs doublets, $B_{\mu\nu}$ and $W^{I}_{\mu\nu}$ are the field strengths of $U(1)_Y$ and $SU(2)_L$, respectively, and $\tau^I = \sigma_I/2$ are the generators of $SU(2)_L$, with $\sigma_I$ being the Pauli matrices. In Eq.~\eqref{OP:SMEFT$+N_R$-dim6}, we additionally have $\tilde{H} = i\sigma_2 H^*$. The operator $\mathcal{O}_{\s NN\gamma}^{ij}$ arises from the $N_R$SMEFT operator at $d = 5$, while the operator $\mathcal{O}_{\s \nu N\gamma}^{\alpha i}$ arises from those at $d = 6$ after expanding around the Higgs vacuum expectation value as $H = (\varphi^+, (v + h + i \varphi_Z)/\sqrt{2})^T$ and transforming to the weak-rotated basis of gauge fields. At the next lowest dimension in the ($N_R$)SMEFT are the $d = 7$ operators~\cite{Fridell:2023rtr}
\begin{align}
\mathcal{O}_{\s LHB}^{(7)\alpha\beta} = (\bar{L}_\alpha\tilde{H})\sigma_{\mu\nu}(\tilde{H}^T L_\beta^c)B^{\mu\nu}\,&,\quad \mathcal{O}_{\s LHW}^{(7)\alpha\beta} = (\bar{L}_\alpha\tilde{H})\sigma_{\mu\nu}(\tilde{H}^T \tau^I L_\beta^c)W^{I\mu\nu}\,, \\
\mathcal{O}_{\s NHB}^{(7)ij} = (\bar{N}_{Ri}^c\sigma_{\mu\nu}N_{Rj})(H^\dagger H)B^{\mu\nu} \,&, \quad \mathcal{O}_{\s NHW}^{(7)ij} =(\bar{N}_{Ri}^c\sigma_{\mu\nu}N_{Rj})(H^\dagger\tau^I H)W^{I\mu\nu} \,,
\end{align}
which generate active-to-active neutrino magnetic moments $\mathcal{O}_{\s \nu\nu\gamma}^{\alpha\beta}$ and further modify $\mathcal{O}_{\s NN\gamma}^{ij}$.

In the context of high-energy collider processes, it is convenient to define a basis of rotated $N_R$SMEFT operators at the EW symmetry breaking scale. In this basis, there are two additional $d = 5$ operators that involve the $Z$ boson field strength,
\begin{align}
\label{OP:ZEFT}
\mathcal{O}_{\s NNZ}^{ij}=(\bar{N}_{Ri}^c\sigma_{\mu\nu}N_{Rj})Z^{\mu\nu} \,,\quad
\mathcal{O}_{\s \nu NZ}^{\alpha i}=(\bar{\nu}_{L\alpha}\sigma_{\mu\nu}N_{Ri})Z^{\mu\nu}\;.
\end{align}
These operators are not present in the usual $N_R$LEFT because $Z$ is integrated out~\cite{Liao:2016qyd,Li:2020lba,Li:2021tsq}; however, they are convenient to use near the $Z$ pole, where they describe the production and subsequent decay of the sterile neutrinos via a dipole-like coupling to the $Z$ boson. We will denote the weak-rotated interactions near the $Z$ pole as
\begin{align}
\label{L:ZEFT}
\mathcal{L}^{\text{$Z$-pole}}_{N_R\text{LEFT}} &\supset d_{\s NN\gamma}^{ij}\, \mathcal{O}_{\s NN\gamma}^{ij} + d_{\s \nu N\gamma}^{\alpha i}\, \mathcal{O}_{\s \nu N\gamma}^{\alpha i} + d_{\s NNZ}^{ij}\, \mathcal{O}_{\s NNZ}^{ij} + d_{\s \nu NZ}^{\alpha i}\, \mathcal{O}_{\s \nu NZ}^{\alpha i} + \text{h.c.}\,.
\end{align}
Now, considering the $N_R$SMEFT operators up to $d = 6$,
the dipole moments in Eq.~(\ref{L:ZEFT}) are related to the $N_R$SMEFT Wilson coefficients in the unbroken phase as
\begin{align}
\label{OP:ZpoleET}
d_{\s{NN\gamma}}^{ij} = c_wC_{\s NNB}^{(5)ij}\,&, \quad d_{\s NNZ}^{ij} = - s_wC_{\s NNB}^{(5)ij} \,,\\
d_{\s{\nu N \gamma}}^{\alpha i} =  \frac{v}{\sqrt{2}} \left(c_w C_{\s{NB}}^{(6)\alpha i} + \frac{s_w}{2} C_{\s{NW}}^{(6)\alpha i}\right)\,&, \quad d_{\s \nu N Z}^{\alpha i} = \frac{v}{\sqrt{2}}\left(- s_w C_{\s NB}^{(6)\alpha i} + \frac{c_w}{2} C_{\s NW}^{(6)\alpha i}\right)\,.
\label{OP:ZpoleET_b}
\end{align}
where $s_w = \sin \theta_w$ and $c_w = \cos\theta_w$, with $\theta_w$ the weak mixing angle, and $v$ is the Higgs vacuum expectation value. From Eq.~\eqref{OP:ZpoleET}, we observe the ratio $d_{\s NNZ}^{\alpha i}/d_{\s NN\gamma}^{\alpha i} = - t_w$ between the sterile-to-sterile couplings, with $t_w = s_w/c_w$. In general, a UV-complete model will also predict some relation between the Wilson coefficients $C_{\s NB}^{(6)\alpha i}$ and $C_{\s NW}^{(6)\alpha i}$; if we take $C_{\s NW}^{(6)\alpha i}/C_{\s NB}^{(6)\alpha i} = a (g/g') = a/t_w$, where $a$ is an arbitrary parameter and $g$ and $g'$ are the $SU(2)_L$ and $U(1)_Y$ couplings, respectively, we obtain the following ratio between the active-to-sterile couplings
\begin{align}
\label{eq:coeff_ratio}
\frac{d_{\s \nu N Z}^{\alpha i}}{d_{\s \nu N \gamma}^{\alpha i}} = \frac{a - 2 t_w^2}{(2 + a)t_w}\,.
\end{align}
There are therefore two flat directions in the parameter space of couplings; $d_{\s \nu N \gamma}^{\alpha i} = 0$ for $a = -2$ and $d_{\s \nu NZ}^{\alpha i} = 0$ for $a = 2t_w^2\approx 0.6$.

In addition to the operators in Eq.~\eqref{OP:ZEFT}, there is also the $d = 5$ charged-current dipole operator,
\begin{align}{\label{L:ZEFTW}}
\mathcal{L}^{\text{$Z$-pole}}_{N_R\text{LEFT}} &\supset d_{\s \ell NW}^{\alpha i}\left(\bar{\ell}_{L\alpha} \sigma_{\mu\nu} N_{Ri} \right)W^{\mu\nu} + \text{h.c.}\,,
\end{align}
where $W^{\mu\nu}$ is the field strength of the $W$ boson. The associated dipole moment is related to the $N_R$SMEFT Wilson coefficient $C_{\s NW}^{(6)}$ in the unbroken phase as
\begin{align}
\label{OP:ZpoleET_W}
d_{\s \ell N W}^{\alpha i} = \frac{v}{2} C_{\s NW}^{(6)\alpha i}\quad \Rightarrow \quad \frac{d_{\s \ell N W}^{\alpha i}}{d_{\s \nu N \gamma}^{\alpha i}} = \frac{\sqrt{2}a}{(2 + a)s_w}\,.
\end{align}
Clearly, we have $d_{\s \ell N W}^{\alpha i} = 0$ for $a = 0$. This operator is important for the single production of a heavy $N$ with a charged lepton or in the decay of $N$ to a charged lepton plus di-jet final state. 

We note that, from a phenomenological point of view, the free parameters of the model can be taken to be $\{m_N, C_{\s NNB}^{(5)ij}, C_{\s NB}^{(6)\alpha i}, C_{\s NW}^{(6)\alpha i}\}$ or $\{m_N, d_{\s N N \gamma}^{ij}, d_{\s \nu N \gamma}^{\alpha i}, a\}$, with general flavour structure. In any UV-complete realisation of these couplings, however, the flavour structure will be fixed. We take both approaches in this paper. First, we consider model-independent constraints on the couplings from LLP searches at the HL-LHC. We then express $d_{\s N N \gamma}^{ij}$ and $d_{\s \nu N \gamma}^{\alpha i}$ in terms of the parameters of a specific UV-complete scenario, which fixes both the flavour structure and the parameter $a$. We will review this UV scenario in the next section.

\section{A UV Model for Sterile Neutrino Magnetic Moments}
\label{sec:model}

%
\begin{table}[]
\centering
\renewcommand{\arraystretch}{1.2}
\begin{tabular}{|c|c|c|c|c|c|}
\hline
Field(s) & Irrep & Couplings & $Z_2$ & Tree-level, $d>4$ & One-loop, $d>4$\\ \hline\hline
$N_R$ & $(\mathbf{1},\mathbf{1})_0$ & $Y_\nu$ & $\checkmark$ & -- & -- \\
$E$ & $(\mathbf{1},\mathbf{1})_{-1}$ & $Y_E$ & $\times$ & $\mathcal{O}_{Hl}^{(1)}$, $\mathcal{O}_{Hl}^{(3)}$ & $\mathcal{O}_{eB}$, $\mathcal{O}_{eW}$, $\ldots$\\
$\phi$ & $(\mathbf{1},\mathbf{1})_{-1}$ & $f$, $\lambda_{\phi H}$ & $\times$ & $\mathcal{O}_{ll}$ & $\mathcal{O}_{eB}$, $\mathcal{O}_{eW}$, $\ldots$ \\\hline
$N_R, \phi$ & -- & $f'$ & $\times$ & $\mathcal{O}_{lNle}$, $\mathcal{O}_{eN}$ & $\mathcal{O}_{\s NB}^{(6)}$, $\mathcal{O}_{\s NW}^{(6)}$, $\dots$ \\
$N_R, E$ & -- & -- & -- & -- & $\mathcal{O}_{\s NB}^{(6)}$, $\mathcal{O}_{\s NW}^{(6)}$, $\dots$ \\
$N_R, E, \phi$ & -- & $h$, $h'$ & $\checkmark$ & -- & $\mathcal{O}_{\s NNB}^{(5)}$, $\ldots$ \\\hline
\end{tabular}
\caption{Fields in the UV model that generate sterile-to-sterile and active-to-sterile neutrino magnetic moments. 
Shown are the representations of the fields under the SM gauge group and the renormalisable couplings when one, two or all of the fields are present; check marks indicate the allowed couplings when $E$ and $\phi$ are odd under $Z_2$. Also shown are a selection of ($N_R$)SMEFT operators generated at tree-level and one-loop after integrating out $E$ and $\phi$, giving interesting observable effects at low energies.}
\label{tab:model}
\end{table}

In this work, we consider a simple UV extension of the SM with RH gauge-singlet fermions $N_R$. To generate magnetic moments for these states, we add the same minimal ingredients as~\cite{Aparici:2009oua,Chala:2020vqp}; a single vector-like lepton $E$ and a scalar $\phi$.\footnote{Comparing to other notation used in the literature, $E \Leftrightarrow X_N^c$~\cite{Chala:2020vqp} and $\phi\Leftrightarrow \varphi$~\cite{Chala:2020vqp}, $h$~\cite{Felkl:2021qdn} and $\mathcal{S}_1^*$~\cite{deBlas:2013gla,Beltran:2023ymm}.} Both are singlets under $SU(3)_c$ and $SU(2)_L$ and have the hypercharges\footnote{We employ the convention $D_\mu = \partial_\mu + ig\tau^I W^{I\mu} + i g' Y B_\mu$ for the covariant derivative, such that SM fields have the hypercharges $Y(L) = -\frac{1}{2}$, $Y(\ell_R) = -1$, $Y(H) = \frac{1}{2}$.} $Y(E) = Y(\phi) = -1$. However, we do not impose an additional $Z_2$ symmetry under which the new fields transform as $E \to -E$ and $\phi \to -\phi$~\cite{Aparici:2009oua}. Without this extra discrete symmetry, the following renormalisable terms can be written, which are not present in the SM,
\begin{align}
\label{eq:model}
\mathcal{L} &\supset \bar{E}\left(i\slashed{D} - m_E\right)E + (D_\mu \phi)^* (D^\mu \phi) - m_\phi^2 |\phi|^2 - \lambda_{\phi\phi} |\phi|^4 - \lambda_{\phi H} |\phi|^2(H^\dagger H)\nonumber \\
&\quad - \Big[\bar{N}_R h E_L \phi^* + \bar{N}_R^c h' E_R \phi^* \nonumber\\ 
&\quad\quad~~ + \bar{L}Y_E H E_R + \bar{E}_L m_{e E} \ell_R + \bar{L}f\tilde{L}\phi + \bar{N}_R^cf'\ell_R \phi^* + \text{h.c.} \Big]\,,
\end{align}
where $\tilde{L} = i\sigma_2 L^c$. The RH neutrinos $N_R$ can also have the Yukawa-type and Majorana mass terms $-\bar{L}Y_\nu \tilde{H} N_R -\frac{1}{2}\bar{N}^c_{R} M_R N_R + \text{h.c.}$, which will be discussed in Section~\ref{sec:neutrino_masses}. Note that, because of the removal of the $Z_2$ symmetry, we obtain four additional terms in the last line of Eq.~\eqref{eq:model} with respect to~\cite{Aparici:2009oua}. The field content of the model is also summarised in Tab.~\ref{tab:model}, where we show the representations of the fields under the SM gauge group and the renormalisable couplings which are present when one, two or all of the fields are considered. With check marks (crosses), we indicate which couplings are allowed (forbidden) when the $Z_2$ symmetry, under which $E$ and $\phi$ are odd, is enforced. We also show the ($N_R$)SMEFT operators which are generated after integrating out $E$ and $\phi$. Among these are the operators $\mathcal{O}_{\s NNB}^{(5)}$, $\mathcal{O}_{\s NB}^{(6)}$ and $\mathcal{O}_{\s NW}^{(6)}$ which induce sterile-to-sterile and active-to-sterile neutrino magnetic moments, but also operators which are relevant for other low-energy probes, such as cLFV processes. 

After EW symmetry breaking, mixing is induced between the SM charged leptons $\ell$ and the vector-like lepton $E$. In the broken phase, the Lagrangian contains the following extended $4\times 4$ mass matrix
\begin{align}
\mathcal{L} \supset 
- \begin{pmatrix}
\bar{\ell}_L & \bar{E}_L
\end{pmatrix}
\mathcal{M}_E
\begin{pmatrix}
\ell_R \\ E_R
\end{pmatrix} + \text{h.c.}\,;\quad \mathcal{M}_E = \begin{pmatrix}
\frac{vY_e}{\sqrt{2}} & \frac{vY_E}{\sqrt{2}}  \\
m_{e E} & m_E
\end{pmatrix}\,,
\end{align}
where the upper-left entry arises from the SM lepton Yukawa term $-\bar{L} Y_e H\ell_R + \text{h.c.}$. Without loss of generality, the weak eigenbasis RH fields $\ell_R$ and $E_R$, which have identical gauge interactions, can be chosen such that all of the entries of $m_{e E}$, which is a $3\times 1$ matrix, are zero~\cite{Alves:2023ufm}. This is equivalent to rotating away non-zero $m_{e E}$ via redefinitions of $Y_e$, $Y_E$ and $m_E$. Now, the mass matrix $\mathcal{M}_E$ can be diagonalised with the following rotation of fields,
\begin{align}
\label{eq:lepton_mixing}
\begin{pmatrix}
    \ell_{L\alpha} \\
    E_L
\end{pmatrix} = \begin{pmatrix}
    V_{\alpha \beta}^L & V_{\alpha E}^L\\
    V_{E \beta}^L & V_{EE}^L
\end{pmatrix} P_L \begin{pmatrix}
    \ell_\beta' \\
    E'
\end{pmatrix} \,,\quad \begin{pmatrix}
    \ell_{R\alpha} \\
    E_R
\end{pmatrix} = \begin{pmatrix}
    V_{\alpha \beta}^R & V_{\alpha E}^R\\
    V_{E \beta}^R & V_{EE}^R
\end{pmatrix} P_R \begin{pmatrix}
    \ell_\beta' \\
    E'
\end{pmatrix}\,,
\end{align}
such that $V^{L\dagger} \mathcal{M}_E V^R = \mathcal{M}_E^{\text{diag}} = \text{diag}(m_e, m_\mu, m_\tau, m_{E'})$. Here, $\ell_\beta'$ and $E'$ represent the physical mass eigenstate fields, with the former corresponding to the known charged leptons $\ell_\beta' \in \{e, \mu, \tau\}$ and the latter a single heavy charged lepton. With abuse of notation, we relabel $\ell_\alpha' \to \ell_\alpha$ and $E' \to E$ in the following. In the limit $Y_e \sim Y_E \ll \sqrt{2}m_E/v$, it becomes possible to block-diagonalise $\mathcal{M}_E$ with the approximate seesaw-like mixing
\begin{align}
\label{eq:seesaw_mixing}
V^L_{\alpha E} = - V_{E \alpha}^{L*} \approx  \frac{v Y_E^{\alpha}}{\sqrt{2}m_E}\,,\quad V^R_{\alpha E} = - V_{E \alpha}^{R*} \approx \frac{v^2 [Y_{e}]_{\alpha\gamma}^*Y_E^{\gamma}}{2 m_E^2}\,,
\end{align}
where we see that the mixing between the RH fields is further suppressed by $v/m_E$ with respect to the mixing between LH fields. Given that the LH mixing is less suppressed, we can also consider the following non-unitarity in the $V^{L}_{\alpha i}$ entry,
\begin{align}
\label{eq:seesaw_mixing_diagonal}
V^L_{\alpha \beta} \approx \delta_{\alpha \beta} - \frac{v^2 Y_E^\alpha Y_E^{\beta *}}{4 m_E^2}\,.
\end{align}
Similarly, non-unitarity is seen in the $V^L_{EE}$ entry, but is irrelevant phenomenologically. 

Expanding the covariant derivatives in the SM Lagrangian and Eq.~\eqref{eq:model}, and transforming to the weak-rotated gauge fields, the gauge interactions of $\nu_L$, $\ell_{L/R}$, $E_{L/R}$ and $\phi$ in the broken phase are found to be
\begin{align}
\label{eq:L_gauge}
\mathcal{L} &\supset - e Q\Big[\bar{\ell}\gamma_\mu\ell + \bar{E}\gamma_\mu E + i\phi^*\overset{\leftrightarrow}{\partial}_\mu\phi\Big]A^\mu -\bigg[\frac{g}{\sqrt{2}}\bar{\ell}_{L}\gamma_\mu \nu_L W^\mu + \text{h.c.}\bigg] \nonumber\\
&\quad~ - \frac{g}{c_w}\bigg[g_L^\nu \bar{\nu}_L\gamma_\mu\nu_L+ g_L^\ell \bar{\ell}_L\gamma_\mu\ell_L 
+ g_R^\ell \bigg(\bar{E}_L\gamma_\mu E_L
+
\begin{pmatrix}
\bar{\ell}_R & \bar{E}_R
\end{pmatrix}
\gamma_\mu \begin{pmatrix}
\ell_R \\ 
E_R
\end{pmatrix} + i\phi^*\overset{\leftrightarrow}{\partial}_\mu\phi\bigg)\bigg]Z^\mu\,,
\end{align}
where, as usual, the electric charge is $Q = \tau^3 + Y$ and the LH and RH $Z$ couplings are $g_L^f = \tau^3 - Q s_w^2$ and $g_R^f = - Q s_w^2$, respectively.
Additionally, the charged lepton and vector-like lepton fields must be rotated to the mass basis according to Eq.~\eqref{eq:lepton_mixing}; the structure of Eq.~\eqref{eq:L_gauge} results in the photon couplings and $Z$ couplings to $\ell_R$ and $E_R$ being diagonal in the mass basis, as insertions of the LH and RH mixing cancel due to unitarity. However, the LH mixing $V^L$ remains in the charged-current term and the neutral-current terms involving $\ell_L$ and $E_L$. In the seesaw limit, the Lagrangian then becomes
\begin{align}
\label{eq:off-diag-Z-couplings}
\mathcal{L} &\supset -\bigg[\frac{g}{\sqrt{2}}\bar{\ell}_{\alpha}\bigg(\delta_{\alpha\beta} - \frac{v^2 Y_{E}^\alpha Y_E^{\beta*}}{4m_E^2}\bigg)\gamma_\mu P_L \nu_\beta W^\mu + \text{h.c.}\bigg] \nonumber \\ 
&\quad\quad - \frac{g}{c_w}\bar{\ell}_\alpha\bigg(g_L^\ell \delta_{\alpha\beta} + \frac{v^2 Y_{E}^\alpha Y_E^{\beta*}}{4m_E^2}\bigg)\gamma_\mu P_L \ell_\beta Z^\mu\,.
\end{align}
The former induces lepton flavour universality (LFU) violating charged-current processes. The latter describes off-diagonal $Z$ couplings or flavour-changing neutral-currents (FCNCs), which are subject to a plethora of stringent bounds from EW precision observables and cLFV. We note that Eq.~\eqref{eq:off-diag-Z-couplings} is equivalent to integrating out $E$ at low energies, which gives non-zero tree-level matching conditions for the coefficients of the $d = 6$ SMEFT operators $\mathcal{O}_{Hl}^{(1)} = (H^\dagger i \overset{\leftrightarrow}{D}_\mu H)(\bar{L}\gamma^\mu L)$ and $\mathcal{O}_{Hl}^{(3)} = (H^\dagger i \overset{\leftrightarrow}{D}\hspace{-0.85em}\phantom{D}^I_\mu H)(\bar{L} \tau^I\gamma^\mu L)$, i.e.,
\begin{align}
C_{Hl}^{(1)\alpha\beta} = C_{Hl}^{(3)\alpha\beta} =- \frac{Y_E^\alpha Y_E^{\beta *}}{4m_E^2} \,,
\end{align}
which has been used previously in the literature~\cite{delAguila:2008pw,deBlas:2017xtg,Crivellin:2020ebi}. Constraints on the Yukawa coupling $Y_E$ from these probes are detailed in Section~\ref{sec:constraints}.

Next, the Lagrangian in Eq.~\eqref{eq:model} results in the Higgs interactions
\begin{align}
\label{eq:L_Higgs}
\mathcal{L} \supset 
- \frac{1}{\sqrt{2}}
\,\bar{\ell}_L
\begin{pmatrix}
Y_e & Y_E
\end{pmatrix}
\begin{pmatrix}
\ell_R \\ E_R
\end{pmatrix}h + \text{h.c.}\,.
\end{align}
This term can also be rotated to the mass basis, where the RH mixing $V^R$ is no longer cancelled due to unitarity. The resulting couplings of the Higgs to charged leptons are
\begin{align}
\label{eq:off-diag_Higgs}
\mathcal{L}\supset - \frac{1}{\sqrt{2}}\bar{\ell}_\alpha\bigg(\delta_{\alpha\gamma} - \frac{3v^2 Y_{E}^\alpha Y_E^{\gamma*}}{4m_E^2}\bigg)[Y_e]_{\gamma\beta} P_R \ell_\beta h + \text{h.c.}\,,
\end{align}
which is nothing other than the tree-level matching of the UV theory onto the $d = 6$ SMEFT operator $\mathcal{O}_{eH} = (H^\dagger H)\bar{L}H\ell_R$, with the coefficient
\begin{align}
C_{eH}^{\alpha\beta} = \frac{Y_{E}^\alpha Y_E^{\gamma*}[Y_e]_{\gamma\beta}}{2m_E^2}\,.
\end{align}
Eq.~\eqref{eq:off-diag_Higgs} modifies SM Higgs decays and induces cLFV Higgs decays via the off-diagonal couplings.

Moving now to the scalar sector of the UV theory, the term $-\bar{L}f\tilde{L}\phi$ in Eq.~\eqref{eq:model} can be written explicitly as
\begin{align}
\label{eq:singly_charged}
\mathcal{L} \supset -
\begin{pmatrix}
\bar{\nu}_L & \bar{\ell}_L
\end{pmatrix}f
\begin{pmatrix}
\ell_L^{c} \\ 
- \nu_L^{c}
\end{pmatrix}\phi
+ \text{h.c.} = -2\bar{\nu}_L f \ell_L^c \phi + \text{h.c.}\,,
\end{align}
where, in the second equality, we have used that $f$ is antisymmetric in flavour space, $f_{\alpha\beta} = - f_{\beta\alpha}$. This interaction also induces LFU violating and cLFV processes at tree-level and one-loop. If the scalar $\phi$ is heavy and integrated out at low energies, Eq.~\eqref{eq:singly_charged} results in the tree-level matching to the SMEFT operator $\mathcal{O}_{ll} = (\bar{L}\gamma_\mu L)(\bar{L}\gamma^\mu L)$, with the coefficient~\cite{deBlas:2014mba,Crivellin:2020klg,Felkl:2021qdn},  
\begin{align}
C_{ll}^{\alpha\beta\gamma\delta} = \frac{f_{\alpha\gamma}f_{\delta \beta}^*}{m_\phi^2}\,.
\end{align}
Including also the term $\bar{N}_R^c f' \ell_R \phi^*$, integrating out $\phi$ also gives tree-level matching to the $N_R$SMEFT operators $\mathcal{O}_{lNle} = (\bar{L} N_{R})\epsilon(\bar{L}\ell_R)$ and $\mathcal{O}_{eN} = (\bar{\ell}_{R}\gamma_\mu \ell_{R})(\bar{N}_R\gamma^\mu N_R)$~\cite{Beltran:2023ymm}, with
\begin{align}
C_{lNle}^{\alpha i \beta \gamma} = \frac{2f_{\alpha\beta}f_{i\gamma}'}{m_\phi^2}\,, \quad C_{eN}^{\alpha \beta ij} = \frac{f_{i\alpha}^{\prime *}f_{j\beta}'}{2m_\phi^2}\,.
\end{align}
This exhausts the tree-level phenomenology of the UV model at low energies, which would have been absent if the $Z_2$ symmetry had been imposed. At the one-loop level, many more operators are induced in the ($N_R$)SMEFT, with partial matching results, predominantly in the SMEFT, available in the literature~\cite{Chala:2020vqp,Zhang:2021jdf,Carmona:2021xtq,Fuentes-Martin:2022jrf}. The focus of the remainder of this section is the one-loop matching of the operators $\mathcal{O}_{\s NNB}^{(5)}$, $\mathcal{O}_{\s NB}^{(6)}$ and $\mathcal{O}_{\s NW}^{(6)}$, which induce LLP signatures at the LHC. First, in Section~\ref{sec:neutrino_masses}, we review the generation of neutrino masses.

\subsection{Neutrino Masses}
\label{sec:neutrino_masses}

For sterile-to-sterile neutrino magnetic moments to be present, at least two RH neutrinos $N_R$ are required in the model. As mentioned below Eq.~\eqref{eq:model}, it is possible to write the renormalisable terms $-\bar{L}Y_\nu \tilde{H}N_R - \frac{1}{2}\bar{N}_R^c M_R N_R + \text{h.c.}$ in the Lagrangian. Below the EW scale, these terms lead to mixing between the active neutrinos $\nu_L$ and $N_R$. The Lagrangian contains
\begin{align}
\mathcal{L} \supset 
- \frac{1}{2}\begin{pmatrix}
\bar{\nu}_L & \bar{N}_R^c
\end{pmatrix}
\mathcal{M}_\nu
\begin{pmatrix}
\nu^c_L \\ N_R
\end{pmatrix} + \text{h.c.}\,;\quad \mathcal{\mathcal{M}}_\nu = \begin{pmatrix}
0 & \frac{vY_\nu}{\sqrt{2}}  \\
\frac{vY_\nu^T}{\sqrt{2}} & M_R
\end{pmatrix}\,,
\end{align}
where the extended neutrino mass matrix $\mathcal{M}_\nu$ is symmetric. Without loss of generality, $M_R$ can be made to be diagonal via a rotation among the RH neutrinos, such that $M_R = \text{diag}(m_{N_1}, m_{N_2},\ldots)$ and the resulting massive states are 
Majorana fermions, with $N_i = N_i^c$. However, we will keep $M_R$ general in the following. The mass matrix can be diagonalised with the following unitary rotation~\cite{Schechter:1980gr}
\begin{align}
\begin{pmatrix}
\nu_{L\alpha}^c \\
N_{Rj}
\end{pmatrix} = \begin{pmatrix}
U_{\alpha i}^* & V_{\alpha N_i}^*\\
\tilde{V}_{ji} & \tilde{U}_{jN_i}
\end{pmatrix} P_R \begin{pmatrix}
\nu_i' \\
N_i'
\end{pmatrix}\,,
\end{align}
such that $V^T \mathcal{M}_\nu V = \mathcal{M}_{\nu}^{\text{diag}} = \text{diag}(m_1,m_2,m_3,m_{N_1'},\ldots)$. Here, $\nu_i' = \nu_i^{\prime c}$ and $N'_i = N_{i}^{\prime c}$ are the physical Majorana mass eigenstate fields in the broken phase, with the former corresponding to the observed light neutrinos and the latter to heavy sterile neutrinos. Similar to the charged leptons, we will relabel $\nu_i' \to \nu_i$ and $N_i' \to N_i$. In the limit $Y_\nu \ll \sqrt{2}M_R/v$, $\mathcal{M}_\nu$ can be diagonalised with
\begin{align}
\label{eq:neutrino_seesaw}
V_{\alpha N_i} &\approx \frac{v}{\sqrt{2}}\big(Y_\nu M_R^{-1}\big)_{\alpha j} \tilde{U}_{jN_i}^*\,,\quad \tilde{V}_{ji} \approx -\frac{v}{\sqrt{2}}\big(M_R^{-1} Y_\nu^T \big)_{j \alpha} U_{\alpha i}^*\,,
\end{align}
to obtain the sub-blocks
\begin{align}
\label{eq:seesaw_masses}
[M_\nu]_{\alpha\beta} = U_{\alpha i}U_{\beta i}m_i&\approx - \frac{v^2}{2}\big(Y_\nu M_R^{-1}Y_\nu^T\big)_{\alpha\beta}\,,\quad [M_N]_{ij} = \tilde{U}^*_{iN_i}\tilde{U}^*_{jN_i}m_{N_i}\approx [M_R]_{ij}\,,
\end{align}
where $U$ and $\tilde{U}$ diagonalise $M_\nu$ and $M_{N}$, respectively. Note that $\tilde{U}_{j N_i} = \delta_{ij}$ if we had taken $M_{R}$ to be diagonal. Combining Eqs.~\eqref{eq:neutrino_seesaw} and \eqref{eq:seesaw_masses}, it is possible to find the well-known relation for the active-sterile neutrino mixing~\cite{Casas:2001sr}
\begin{align}
\label{eq:casas-ibarra}
V_{\alpha N_i}  = i  U_{\alpha j} \mathcal{R}_{ji} \sqrt{\frac{m_{j}}{m_{N_i}}} \,,
\end{align}
where $\mathcal{R}$ is an arbitrary orthogonal matrix, i.e. $\mathcal{R}_{ji}\mathcal{R}_{ki} = \delta_{jk}$.

In this work, we consider heavy sterile states $N_i$ within the kinematic reach of the HL-LHC, i.e., $m_{N_i}$ from 1~GeV up to 1~TeV. For this range of masses, there are two possible ways to produce the observed light neutrino masses via the tree-level relations Eq.~\eqref{eq:seesaw_masses}. The first scenario is that the heavy states $N_i$ are dominantly Majorana (or, in other words, the masses $m_{N_i}$ are well separated). Then, the light neutrino masses are given by $[M_\nu]_{\alpha\beta} = -\frac{v^2}{2} [Y_\nu]_{\alpha i}[Y_\nu]_{\beta i}/m_{N_i}$, which is the standard Type-I seesaw relation~\cite{Minkowski:1977sc,Gell-Mann:1979vob,Yanagida:1979as,Mohapatra:1979ia,Schechter:1980gr}. For $m_{N} \sim 100$~GeV, neutrino masses $m_\nu \sim 0.1$~eV then require Yukawa couplings of size $[Y_{\nu}]_{\alpha i} \sim 10^{-6}$. The active-sterile mixing is governed by Eq.~\eqref{eq:casas-ibarra} with $|\mathcal{R}_{ji}|\leq 1$, or $V_{\ell N}\sim \sqrt{m_\nu/m_N}$. However, for heavy sterile states in the mass range relevant for the HL-LHC, this active-sterile mixing $V_{\ell N}$ is too suppressed to give observable effects unless some additional NP is present. 

The second scenario, which can be more relevant for LLP searches, is when pairs of heavy sterile states $N_i$ form pseudo-Dirac states with small mass splittings. In this limit, cancellations in the matrix product $Y_\nu M_R^{-1} Y_\nu^T$ ensure small light neutrino masses instead of suppressed Yukawa couplings. Now that $Y_\nu$ can in principle be large, so can the active-sterile mixing in Eq.~\eqref{eq:neutrino_seesaw}; equivalently, this corresponds to the limit $|\mathcal{R}_{ji}| \gg 1$ in Eq.~\eqref{eq:casas-ibarra}. This interesting scenario is minimally obtained in the inverse seesaw mechanism~\cite{Mohapatra:1986aw,Nandi:1985uh,Mohapatra:1986bd,Gonzalez-Garcia:1988okv,Deppisch:2004fa,Ma:2009du,Dev:2012sg,BhupalDev:2012jvh,CentellesChulia:2020dfh}, which does not require any additional Higgs doublets or symmetries. 
To demonstrate this mechanism, we can consider the simplified scenario where two heavy sterile states, $N_{R1} \equiv N_R$ and $N_{R2} \equiv S_L^c$,
only interact with one generation of active neutrino. The relevant mass terms can be written as
\begin{align}
\label{massterm}
\mathcal{L} \supset - m_{\nu N} \bar{\nu}_L N_R - m_{\nu S} \bar{\nu}_L S_L^{c} - {1\over 2 } \left(\mu_N \bar{N}_R^c N_R + \mu_{S}  \bar{S}_L S_L^{c} \right) - m_{N_D} \bar{S}_L N_R + {\rm h.c.}\,,
\end{align}
where $(m_{\nu N}, m_{\nu S}) \equiv v Y_\nu/\sqrt{2}$. As such, the extended neutrino mass matrix (assuming one generation of active neutrino) in the basis $(\nu_L^c, N_R, S^{c}_L)$ takes the form
\begin{align}
\label{eq:extended_mass_matrix}
\mathcal{M}_\nu = \left( \begin{matrix}  0 &  m_{\nu N} & m_{\nu S} \cr  m_{\nu N} & \mu_N & m_{N_D} \cr m_{\nu S}  & m_{N_D} & \mu_{S}\end{matrix} \right) \,, 
\end{align}
where $m_{\nu S}$ can be rotated away without any loss of generality~\cite{Ma:2009du,Abada:2014vea} and the lepton number violating masses $\mu_{N,S}\ll m_{N_D}$ are naturally small in the sense of 't~Hooft~\cite{tHooft:1979rat}, i.e. total lepton number symmetry is restored in the limit $ \mu_{N,S}\to 0$. The symmetric mass matrix $\mathcal{M}_\nu$ can be diagonalised as before; the inclusion of $N_R$ and $S_L^c$ leads to two Majorana eigenstates $N_1$ and $N_2$ with a small mass splitting compared to the Dirac mass $m_{N_D}$. The unitary rotation matrix $V$ can be solved perturbatively by using the Hermitian combination $\mathcal{M}_\nu^\dagger \mathcal{M}_\nu$ (or $\mathcal{M}_\nu \mathcal{M}_\nu^\dagger$),
\begin{align}
\label{eq:diagsq}
V^\dagger \mathcal{M}^\dagger_\nu \mathcal{M}_\nu V
= \left(V^T \mathcal{M}_\nu V \right)^\dagger \left( V^T \mathcal{M}_\nu V \right) = \text{diag}(m_{\nu}^2,m_{N_1}^2,m_{N_2}^2)
\,,
\end{align}
where the $V$ diagonalising $\mathcal{M}_\nu \mathcal{M}_\nu^\dagger$ is the same as that in Eq.~\eqref{eq:diagsq}. The most straightforward case is obtained in the limit where $\mu_N,\mu_{S}\ll m_{\nu N}\ll m_{N_D}$ and, without loss of generality, $m_{\nu S}$ is set to zero via a rotation among the sterile states. In this limit, the mass eigenvalues can be solved perturbatively to obtain
\begin{align}
\label{eq:eigensq-1}
m_\nu &\simeq \frac{|m_{\nu N}|^2 |\mu_{S}|}{|m_{\nu N}|^2+|m_{N_D}|^2} \,, \\
m_{N_1}^2 &\simeq |m_{\nu N}|^2 + |m_{N_D}|^2 -\frac{\left|\mu_{S}^* m_{N_D}^2+ \mu_N\left( |m_{\nu N}|^2 + |m_{N_D}|^2\right)\right|}{\sqrt{|m_{\nu N}|^2+|m_{N_D}|^2}}\,, \\
m_{N_2}^2 &\simeq |m_{\nu N}|^2 + |m_{N_D}|^2 + \frac{\left|\mu_{S}^* m_{N_D}^2+ \mu_N \left(|m_{\nu N}|^2 + |m_{N_D}|^2\right)\right|}{\sqrt{|m_{\nu N}|^2+|m_{N_D}|^2}}\,.
\end{align}
Another interesting case occurs when $\mu_N \sim \mu_{S}$ and $m_{\nu N}\sim m_{\nu S}$, requiring some additional symmetry preventing the freedom to rotate away $m_{\nu S}$. 
The mass matrix can be exactly solved in this scenario, and assuming the parameters to be real, it leads to the masses
\begin{align}
\label{eq:eigensq-2}
m_\nu^2 &= 2 m_{\nu N}^2 + \frac{1}{2} (m_{N_D}+\mu_S)^2\left[ 1-\sqrt{1+\frac{8 m_{\nu N}^2}{(m_{N_D}+\mu_S)^2}}\right] \,, \\
m_{N_1}^2 &= 2 m_{\nu N}^2 + \frac{1}{2} (m_{N_D}+\mu_S)^2\left[1+\sqrt{1+\frac{8 m_{\nu N}^2}{(m_{N_D}+\mu_S)^2}}\right] \,, \\
m_{N_2}^2 &= (m_{N_D} - \mu_S)^2\,.
\end{align}
This case corresponds to the scenario where $N_2$ is decoupled, which can be protected in the presence of an explicit symmetry, and only $N_1$ mixes with the active state. We note that any new interactions involving $N_1$ and $N_2$, e.g., a transition magnetic dipole moment, is in general not diagonal in this basis and may lead to corrections to the mass matrix $\mathcal{M}_\nu$. However, to restrict ourselves to these two limiting cases of the inverse seesaw mechanism, we will assume any such correction to be small.

\begin{figure}[t!]
\centering
\includegraphics[width=0.24\textwidth]{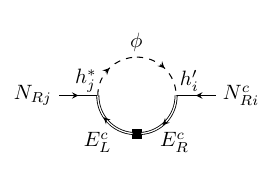}
\includegraphics[width=0.24\textwidth]{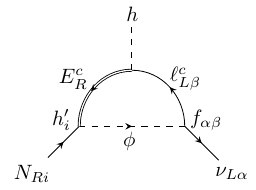}
\caption{Diagrams at one-loop in the UV model which renormalise the RH neutrino mass matrix $M_R$ (left) and Yukawa coupling $Y_\nu$ (right). For the Yukawa coupling, there are two other one-loop diagrams; one similar to that above with the replacements $E_R^c\to \ell^c_{R\rho}$, $h_i' \to f_{i\rho}'$ and $Y_{E}^{\beta*}\to [Y_{e}]_{\beta\rho}^*$ and another with an internal Higgs line and an insertion of $[Y_\nu]_{\alpha i}$.}
\label{fig:nu-asymm4}
\end{figure}

In the presence of the fields $E$ and $\phi$ and their interactions in Eq.~\eqref{eq:model}, it is possible to draw the one-loop diagrams in Fig.~\ref{fig:nu-asymm4} which renormalise the Yukawa coupling $Y_\nu$ and the RH neutrino mass $M_R$. The threshold corrections from these diagrams are
\begin{align}
[Y_{\nu}]_{\alpha i} &= [Y_{\nu}^{(0)}]_{\alpha i} - \frac{1}{8\pi^2}\bigg[f_{\alpha\beta}Y_{E}^{\beta*}h_i'\bigg(1 + \frac{r\log r}{1-r} + \log\frac{\mu^2}{m_E^2} \bigg) + f_{\alpha\beta}[Y_e]_{\beta\rho}^*f_{i\rho}'\bigg(1 + \log\frac{\mu^2}{m_\phi^2} \bigg) \nonumber \\
&\hspace{8em} - \frac{1}{2}Y_E^\alpha Y_E^{\beta*}[Y_\nu^{(0)}]_{\beta i} \bigg(1 - \frac{\mu_H^{(0)2}}{m_E^2} \bigg)\bigg(1 + \log\frac{\mu^2}{m_E^2} \bigg)\bigg]\,, \\
[M_{R}]_{ij} &= [M_{R}^{(0)}]_{ij} - \frac{(h_i'h_j^* + h_i^*h_j')m_E}{16\pi^2}\left(1 + \frac{r\log r}{1-r} + \log\frac{\mu^2}{m_E^2} \right)\,,
\label{eq:MR_renorm}
\end{align}
where $r = m_\phi^2/m_E^2$ and $\mu_H$ is the usual parameter in the Higgs potential. While it is always possible to absorb the finite corrections to $Y_\nu$ and $M_R$ in their renormalised values, these expressions are nevertheless useful to verify if fine-tuning between the tree-level and one-loop contributions is required to obtain the desired heavy masses $m_{N_i}$ and transition magnetic moments, calculated in the following sections.

\subsection{Sterile-to-Sterile Neutrino Magnetic Moments}

\begin{figure}[t!]
\centering
\includegraphics[width=0.24\textwidth]{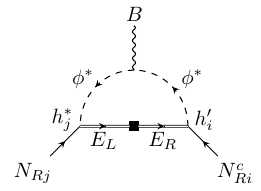}
\includegraphics[width=0.24\textwidth]{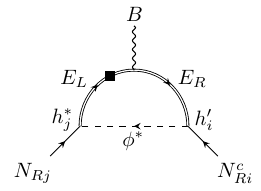}
\includegraphics[width=0.24\textwidth]{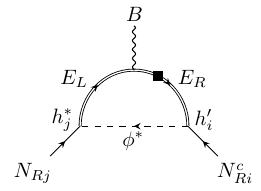}
\caption{Diagrams at one-loop in the UV model 
which generate the $d = 5$ dipole operator $\mathcal{O}_{\s NNB}^{(5)}$. The black square indicates an insertion of the vector-like lepton mass $m_E$. At low energies, these induce electromagnetic sterile-to-sterile dipole moments $\mathcal{O}_{\s NN\gamma}$. Near the $Z$ pole, sterile-to-sterile dipole couplings to $Z$, $\mathcal{O}_{\s NNZ}$, are also induced.}
\label{fig:nu-asymm1}
\end{figure}

In our model, the $d = 5$ $N_R$SMEFT dipole operator $\mathcal{O}_{\s NNB}^{(5)}$ is generated at one-loop, as shown by the diagrams in Fig.~\ref{fig:nu-asymm1}, which then induces sterile-to-sterile transition magnetic moments $\mathcal{O}_{\s NN\gamma}$ and $\mathcal{O}_{\s NNZ}$ in the broken phase. We perform the matching of the UV model to the effective operator coefficient $C_{\s NNB}^{(5)}$ using the diagrammatic approach, which involves computing the one-light-particle-irreducible (1LPI) amplitudes in the effective and UV theories and requiring that they coincide at the matching scale. All one-loop amplitudes are performed using dimensional regularisation with $d = 4-2\epsilon$ space-time dimensions and introducing the renormalisation scale $\mu$. We then subtract divergent pieces in the $\overline{\text{MS}}$ scheme.

The amplitude $\braket{N_{i} N_{j} B}$ in the $N_R$SMEFT via the operator $\mathcal{O}_{\s NNB}^{(5)ij}$ is
\begin{align}
\label{eq:NNB_EFT_amplitude}
i \mathcal{M}_{ij}^{\text{EFT}} &= 4 \bar{u}(p_{N_i}) \sigma_{\mu\nu} p_B^\nu \Big[C_{\s NNB}^{(5)ij} P_R - C_{\s NNB}^{(5)ij*}P_L\Big] u(p_{N_j}) \epsilon^{\mu*}(p_B)\,,
\end{align}
where $p_B = p_{N_j} - p_{N_i}$. Here, the indices $i,j$ are fixed so there is an additional factor of 2 in the amplitude to account for the contribution from the antisymmetric term $-C_{NNB}^{(5)ji}$, which arises because of the Majorana nature of the sterile neutrinos. The amplitude $\braket{N_i N_j B}$ in the UV model is instead given by
\begin{align}
i \mathcal{M}_{ij}^{\text{UV}} &= g' Y \bar{u}(p_{N_i})\Big[(h_i' P_R + h_i P_L) \Gamma_\mu (h_j^* P_R + h_j^{\prime *} P_L) \nonumber \\
&\hspace{6em} - (h_i^{*} P_R + h_i^{\prime *} P_L) \Gamma_\mu (h_j^{\prime} P_R + h_j P_L) \Big]u(p_{N_j})\epsilon^{\mu*}(p_B) \,,
\end{align}
where $Y(E) = Y(\phi) = -1$ and $\Gamma_\mu$ denotes the loop integral
\begin{align}
\label{eq:CNNB_loop-full}
\Gamma_\mu &= \mu^{2\epsilon}\int\frac{d^{d}k}{(2\pi)^d}\,\bigg[\frac{(\slashed{p}\hspace{-0.5em}\phantom{p}_{N_i} -\slashed{k}+ m_E)\gamma_\mu (\slashed{p}\hspace{-0.5em}\phantom{p}_{N_j}-\slashed{k}+ m_E)}{\big[k^2 - m_\phi^2\big]\big[(p_{N_i}-k)^2 - m_E^2\big]\big[(p_{N_j} -k)^2 - m_E^2\big]}\nonumber \\
& \hspace{7.5em}-\frac{(\slashed{k}+ m_E)(p_{N_i} + p_{N_j} - 2k)_\mu}{\big[k^2 - m_E^2\big]\big[(p_{N_i} -k)^2 - m_\phi^2\big]\big[(p_{N_j}-k)^2 - m_\phi^2\big]}\bigg]\,,
\end{align}
where $k$ is the loop momentum. The first and second terms inside the square brackets of Eq.~\eqref{eq:CNNB_loop-full} originate from the coupling of $B_\mu$ to $E$ and $\phi$, respectively. We now expand the loop integral in powers of the external momenta and, assuming $m_{N_{i}}\ll m_E, m_\phi$, obtain the amplitude 
\begin{align}
\label{eq:NNB_UV_amplitude}
i \mathcal{M}_{ij}^{\text{UV}} &= \frac{g'}{16\pi^2 m_E} f(r) \bar{u}(p_{N_i}) \sigma_{\mu\nu} p_B^\nu \Big[(h'_i h^*_j - h^*_i h'_j)P_R - (h^{\prime *}_i h_j - h_i h^{\prime *}_j)P_L\Big] u(p_{N_j})\epsilon^{\mu*}(p_B)\,,
\end{align}
where we define the loop function
\begin{align}
\label{eq:CNNB_loop}
f(r) = \frac{1}{1-r} + \frac{r\log r}{(1-r)^2} \,,
\end{align}
again with $r = m_\phi^2/m_E^2$. For $m_E = m_\phi$, the limit of this function is $f(r)\big|_{r\to 1} = \frac{1}{2}$.

The amplitudes in Eqs.~\eqref{eq:NNB_EFT_amplitude} and~\eqref{eq:NNB_UV_amplitude} can now be equated to find
\begin{align}
\label{eq:CNNB_matching}
C_{\s NNB}^{(5)ij} = \frac{g'(h_i'h_j^* - h_i^* h_j')}{64\pi^2 m_E} f(r) \,,
\end{align}
which is in agreement with the result of~\cite{Aparici:2009oua}. We see that the flavour structure is determined entirely by the combination $h_i'h_j^* - h_i^* h_j'$, with the diagonal elements vanishing, as expected.  In the broken phase of the SM, but at energies where the $Z$ boson is not yet integrated out, ${C}_{\s NNB}^{(5)}$ results in the $\gamma$ and $Z$ couplings ($d_{\s{NN\gamma}}$ and $d_{\s{NNZ}}$) according to Eq.~\eqref{OP:ZpoleET}. For $m_E = m_\phi$, Eqs.~\eqref{eq:CNNB_matching} and \eqref{OP:ZpoleET} give dipole couplings of size,
\begin{align}
d_{\s NN\gamma}^{ij} = c_w C_{\s NNB}^{(5)ij} \approx 2.4 \times 10^{-6}~\text{GeV}^{-1}\bigg(\frac{h_i'h_j^* - h_i^* h_j'}{10}\bigg)\bigg(\frac{1~\text{TeV}}{m_E}\bigg)\,.
\end{align}
Thus, even with values of the couplings at the perturbative limit, $h, h' \lesssim \sqrt{4\pi}$, values of the dipole coupling can only be obtained up to $d_{\s NN\gamma}^{ij}\sim 10^{-4}~\text{GeV}^{-1}$ with $m_E$ and $m_\phi$ just above the current lower limits from collider searches, $m_E, m_\phi \gtrsim 200$~GeV (see Section~\ref{sec:constraints} for further details).

Examining Eq.~\eqref{eq:MR_renorm}, we observe that the one-loop correction to $[M_R]_{ij}$ depends on the same couplings entering $C_{\s NNB}^{(5)ij}$ in Eq.~\eqref{eq:CNNB_matching}. Therefore, to obtain large sterile-to-sterile neutrino magnetic moments for sterile states in the 10~GeV to 1~TeV range, some fine-tuning between the tree-level and one-loop contributions to $[M_R]_{ij}$ may be required. 

\subsection{Active-to-Sterile Neutrino Magnetic Moments}

%
\begin{figure}[t!]
\centering
\includegraphics[width=0.24\textwidth]{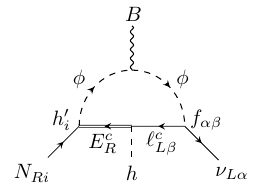}
\includegraphics[width=0.24\textwidth]{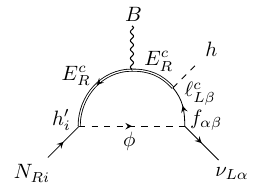}
\includegraphics[width=0.24\textwidth]{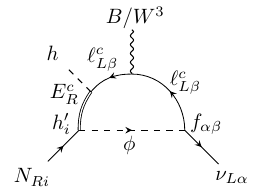}
\caption{Diagrams at one-loop in the UV model which generate the $d = 6$ dipole operators $\mathcal{O}_{\s NB}^{(6)}$ and $\mathcal{O}_{\s NW}^{(6)}$. At low energies, these induce the electromagnetic active-to-sterile dipole moment $\mathcal{O}_{\s \nu N \gamma}$. 
Additionally, at energies near the EW scale, the dipole moments $\mathcal{O}_{\s \ell N W}$ and $\mathcal{O}_{\s \nu N Z}$ are generated. Note that there are additional diagrams with $E_R^c\to \ell^c_{R\rho}$, $h_i' \to f_{i\rho}'$ and $Y_{E}^{\beta*}\to [Y_{e}]_{\beta\rho}^*$ and others with an internal Higgs line and an insertion of $[Y_\nu]_{\alpha i}$, which also contribute to Eq.~\eqref{eq:CNB_matching}.}
\label{fig:nu-asymm2}
\end{figure}

The $d = 6$ operators $\mathcal{O}_{\s NB}^{(6)}$ and $\mathcal{O}_{\s NW}^{(6)}$, which induce active-to-sterile neutrino transition magnetic moments $\mathcal{O}_{\s \nu N\gamma}$ and $\mathcal{O}_{\s \nu NZ}$, as well as the charged-current dipole $\mathcal{O}_{\s \ell NW}$, are also generated in the UV model at one-loop, as depicted in Fig.~\ref{fig:nu-asymm2}. These diagrams do not require mixing between the active and sterile neutrinos and thus avoid mixing suppression. However, they do rely on the Yuwawa coupling $Y_E$ of the lepton doublet $L$ with vector-like lepton $E_R$ and Higgs doublet $H$. In the following, we use the same procedure as the previous section to find the matching between $C_{\s NB}^{(6)}$ and $C_{\s NW}^{(6)}$ and the parameters of the UV model.

In the $N_R$SMEFT, the amplitude $\braket{\nu_{\alpha} N_i B h}$ is given by
\begin{align}
\label{eq:nuNB_EFT_amplitude}
i\mathcal{M}_{\alpha i}^{\text{EFT}} = \sqrt{2} C_{\s NB}^{(6)\alpha i}  \bar{u}(p_{\nu_\alpha}) \sigma_{\mu\nu} p_B^\nu P_Ru(p_{N_i}) \epsilon^{\mu*}(p_B)\,,
\end{align}
where $p_B = p_{N_i} - p_{\nu_\alpha}$. In the UV model, the amplitude $\braket{\nu_\alpha N_i B h}$ is determined in part by the three one-loop diagrams shown in Fig.~\ref{fig:nu-asymm2}. It is also possible to draw one-loop diagrams analogous to those in Fig.~\ref{fig:nu-asymm2}, but with $E_R^c\to \ell^c_{R\rho}$, $h_i' \to f_{i\rho}'$ and $Y_{E}^{\beta*}\to [Y_{e}]_{\beta\rho}^*$, and another with an internal Higgs line proportional to $Y_E^\alpha Y_E^{\beta*}[Y_\nu]_{\beta i}$. Including all of these contributions, the UV amplitude $\braket{\nu_\alpha N_i B h}$ can be calculated and compared with the EFT amplitude, as enacted in the previous section. This gives the matching condition,
\begin{align}
\label{eq:CNB_matching}
C_{\s NB}^{(6)\alpha i} = \frac{g'}{64\pi^2m_E^2}\left[3 f_{\alpha\beta}Y_E^{\beta *} h_i' f(r) - \frac{f_{\alpha\beta}[Y_e]_{\beta\rho}^* f_{i\rho}'}{r} \bigg(\frac{5}{2} + 3\log\frac{\mu^2}{m_\phi^2}\bigg) + Y_E^{\alpha}Y_E^{\beta*}[Y_\nu]_{\beta i}\right] \,,
\end{align}
where the flavour indices $\beta$ and $\rho$ are summed over and the loop function $f(r)$ is the same as in Eq.~\eqref{eq:CNNB_loop}, with $r = m_\phi^2/m_E^2$.

As seen in the diagram to the right of Fig.~\ref{fig:nu-asymm2}, the intermediate charged lepton can also couple to $W_\mu^3$, which generates the operator $\mathcal{O}_{\s NW}^{(6)}$. The amplitude for $\braket{\nu_\alpha N_i W^3 h}$ from this operator is
\begin{align}
\label{eq:nuNW_EFT_amplitude}
i\mathcal{M}_{\alpha i}^{\text{EFT}} = \frac{C_{\s NW}^{(6)\alpha i}}{\sqrt{2}}\bar{u}(p_{\nu_\alpha}) \sigma_{\mu\nu} p_{W}^\nu P_R u(p_{N_i}) \epsilon^{\mu*}(p_{W})\,,
\end{align}
for $p_{W} = p_{N_i} - p_{\nu_\alpha}$. The UV amplitude $\braket{\nu_\alpha N_i W^3 h}$ can be calculated from the diagrams entering $\braket{\nu_\alpha N_i B h}$ where it is possible to replace $B\to W^{3}$. The resulting matching condition is
\begin{align}
\label{eq:CNW_matching}
C_{\s NW}^{(6)\alpha i} = \frac{g}{32\pi^2 m_E^2}\left[f_{\alpha\beta}Y_E^{\beta *} h_i' f(r) - \frac{f_{\alpha\beta}[Y_e]_{\beta\rho}^* f_{i\rho}'}{r}\bigg(\frac{3}{2} + \log \frac{\mu^2}{m_\phi^2}\bigg) + Y_E^{\alpha}Y_E^{\beta*}[Y_\nu]_{\beta i} \right]\,.
\end{align}
From Eqs.~\eqref{eq:CNB_matching} and \eqref{eq:CNW_matching} we see that, in principle, there are multiple contributions to these operators. To simplify this dependence on the UV model parameters, we note that we are only interested in the scenario where the coefficient $C_{\s NNB}^{(5)ij}$ is sizeable, requiring large $h'$. Then, if we assume that the first terms in Eqs.~\eqref{eq:CNB_matching} and \eqref{eq:CNW_matching} dominate over the others, we obtain the particularly simple relation between the coefficients
\begin{align}
\label{eq:CNW_matching-2}
C_{\s NW}^{(6)\alpha i} = \frac{2}{3t_w}C_{\s NB}^{(6)\alpha i}\approx 1.22 \,C_{\s NB}^{(6)\alpha i}\,.
\end{align}
Thus, as discussed in Section~\ref{sec:EFT}, the UV model sets the parameter $a = 2/3$. Now, using Eqs.~\eqref{OP:ZpoleET} and Eq.~\eqref{eq:CNW_matching}, we find the corresponding active-to-sterile dipole couplings in the broken phase to be
\begin{align}
d_{\s \nu N\gamma}^{\alpha i} = \frac{4 v c_w}{3\sqrt{2}}C_{\s NB}^{(6)\alpha i} \approx 1.7 \times 10^{-9}~\text{GeV}^{-1} \bigg(\frac{f_{\alpha\beta}Y_E^{\beta*}h_i'}{10^{-2}}\bigg)\bigg(\frac{1~\text{TeV}}{m_E}\bigg)^2\,.
\label{eq:dnuNratio}
\end{align}
Using Eqs.~\eqref{eq:coeff_ratio} and $a = 2/3$, we also find the ratio $d_{\s \nu N Z}^{\alpha i}/d_{\s \nu N \gamma}^{\alpha i} = (1-3t_w^2)/(4 t_w) \approx 4.5\times 10^{-2}$. The active-to-sterile dipole coupling with $Z$ is therefore relatively suppressed with respect to $d_{\s \nu N \gamma}^{\alpha i}$ in this scenario. Finally, we note that the flavour structure of the couplings $d^{\alpha i}_{\s \nu N \gamma}$ and $d^{\alpha i}_{\s \nu N Z}$ exhibits an interesting dependence on the parameters of the UV model. In the case of flavour universal couplings, at least one of the entries of $d^{\alpha i}_{\s \nu N \gamma}$ and $d^{\alpha i}_{\s \nu N Z}$ for $\alpha \in \{e,\mu,\tau\}$ must vanish, with the other two being equal and opposite in sign; for the choice $Y_E^e = Y_E^\mu = Y_E^\tau$ and $f_{e\mu} = f_{e\tau} = f_{\mu\tau}$, we obtain $d^{\mu i}_{\s \nu N \gamma} = d^{\mu i}_{\s \nu N Z} = 0$, for example. In the scenario where only the $\mu$-$\tau$ couplings are non-zero, for example $Y_E^\mu = Y_E^\tau$ and $f_{\mu\tau}$, we would instead have $d^{e i}_{\s \nu N \gamma} = d^{e i}_{\s \nu N Z} = 0$, $d^{\mu i}_{\s \nu N \gamma} = - d^{
\tau i}_{\s \nu N \gamma}$ and $d^{\mu i}_{\s \nu N Z} = - d^{
\tau i}_{\s \nu N Z}$. Finally, if only couplings involving $\tau$ are present (e.g., $Y_E^\tau$, $f_{e\tau} = f_{\mu\tau}$), we would obtain $d^{\tau i}_{\s \nu N \gamma} = d^{\tau i}_{\s \nu N Z} = 0$, $d^{e i}_{\s \nu N \gamma} = d^{
\mu i}_{\s \nu N \gamma}$ and $d^{e i}_{\s \nu N Z} = d^{
\mu i}_{\s \nu N Z}$. These particular limits can be readily translated to other flavour combinations.

\subsection{Active-to-Active Neutrino Magnetic Moments}

%
\begin{figure}[t!]
\centering
\includegraphics[width=0.24\textwidth]{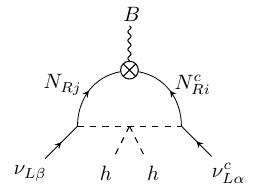}
\includegraphics[width=0.24\textwidth]{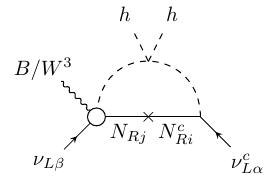}
\includegraphics[width=0.24\textwidth]{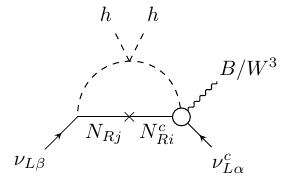}
\caption{Diagrams at two loops in the UV model which contribute negligibly to the $d = 7$ dipole operators $\mathcal{O}_{\s LHB}^{(7)}$ and $\mathcal{O}_{\s LHW}^{(7)}$. The crossed circle in the left diagram refers to the sterile-to-sterile one-loop diagrams in Fig.~\ref{fig:nu-asymm1}, while the circle in the other two diagrams correspond to the active-to-sterile diagrams in Fig.~\ref{fig:nu-asymm2}.}
\label{fig:nu-asymm3}
\end{figure}

Here, we finally comment that active-to-active neutrino magnetic moments are naturally suppressed in the UV model. With the fields and couplings available in Eq.~\eqref{eq:model} and assuming the RH neutrinos satisfy $m_{N_{i}} \ll m_E, m_\phi$, the lowest-order 1LPI amplitudes that contribute to the $d = 7$ SMEFT operators $C_{\s LHB}^{(7)\alpha\beta}$ and $C_{\s LHW}^{(7)\alpha\beta}$ are at two loops, as shown in Fig.~\ref{fig:nu-asymm3}.

However, in the broken phase, and at energies where the RH neutrinos may also be integrated out, active-to-active neutrino magnetic moments are induced via the active-sterile mixing $V_{\ell N}$, i.e.,
\begin{align}
d_{\s \nu\nu\gamma}^{\alpha \beta} &= V_{\alpha N_i}^* V_{\beta N_j}^* d^{ij}_{\s N N \gamma} + \Big[V_{\alpha N_i}^* d_{\s \nu N\gamma}^{\beta i} - V_{\beta N_i}^* d_{\s \nu N\gamma}^{\alpha i}\Big]  \nonumber\\
&\approx 10^{-16}~\text{GeV}^{-1}\bigg[\bigg(\frac{V_{\alpha N_i}^*V_{\beta N_j}^*}{10^{-10}}\bigg)\bigg(\frac{d_{\s NN\gamma}^{ij}}{10^{-6}~\text{GeV}^{-1}}\bigg) + \bigg(\frac{V_{\alpha N_i}^*}{10^{-5}}\bigg)\bigg(\frac{d_{\s \nu N\gamma}^{\beta i}}{10^{-11}~\text{GeV}^{-1}}\bigg)\bigg] \,,
\end{align}
where in the second line we have assumed that $\alpha\neq \beta$ and $d^{\alpha i}_{\s \nu N \gamma} = 0$. For active-sterile mixing of the size predicted by the Type-I seesaw mechanism, $V_{\ell N}\sim \sqrt{m_\nu/m_N}\sim 10^{-6}$, this contribution is highly suppressed both by the mixing and the one-loop suppressed $d_{\s N N \gamma}^{ij}$ and $d_{\s \nu N \gamma}^{\alpha i}$. Thus, even with the larger mixing $V_{\ell N}$ in the inverse seesaw scenario, we can still expect the active-to-active neutrino magnetic moments to safely satisfy the bounds from TEXONO~\cite{TEXONO:2006xds}, GEMMA~\cite{Beda:2007hf}, LSND~\cite{LSND:2001akn} and Borexino~\cite{Borexino:2008dzn}, $d_{\s \nu\nu\gamma}^{\alpha\beta}\lesssim 2\times 10^{-8}~\text{GeV}^{-1}$.

\section{Phenomenological Constraints on the UV Model}
\label{sec:constraints}

In this section, we review the constraints on the UV-complete extension of the SM outlined in Section~\ref{sec:model}. We begin by examining the direct production of $E$ and $\phi$ in collider experiments. We then consider rare low-energy processes which are not present or highly-suppressed in the SM, but could be enhanced by non-diagonal flavour couplings after integrating out $E$ and $\phi$.

\subsection{Constraints from Direct Collider Searches}
\label{subsec:collider}

A recasting of direct searches for selectrons and smuons at the LHC provides useful constraints on the singly-charged scalar $\phi$ in our UV scenario. 
The dominant limit comes from the Drell-Yan pair production process $pp \to \gamma/Z \to\phi^+ \phi^-$ followed by the decay $\phi^{\pm}\rightarrow \ell^{\pm}\nu$. In~\cite{Cao:2017ffm,Alcaide:2017dcx,Alcaide:2019kdr,Crivellin:2020klg}, the most recent ATLAS search~\cite{ATLAS:2019lff} for oppositely-charged electron and muon pairs with 139~$\text{fb}^{-1}$ of collected data was recast for the singly-charged scalar scenario. The ATLAS search places a lower bound on the slepton masses of approximately 450~GeV for a 100\% branching ratio of the slepton in the specific channel. The major difference between this analysis and the pair production of $\phi^+\phi^-$ is the cross-section in the two scenarios. In~\cite{Crivellin:2020klg}, a simple scaling factor was used to map the leading-order Madgraph-simulated production cross-section for $pp \to \phi^+ \phi^-$ onto the production cross-section given by ATLAS for the RH slepton pair. Allowing for further uncertainties, a conservative lower limit of $m_\phi \gtrsim 200$~GeV was found. In addition, monophoton searches for dark matter at LEP~\cite{DELPHI:2003dlq,Fox:2011fx} can also be recast to obtain the limit $m_\phi/|f_{e\mu}| \gtrsim 350~ \text{GeV}$~\cite{Crivellin:2020klg}.

In general, $SU(2)_L$ singlet vector-like leptons $E$ are subject to much weaker bounds compared to their $SU(2)_L$ doublet counterparts, due to their relatively smaller production cross-sections. Initially, searches at LEP~\cite{L3:2001xsz} were able to exclude the masses $m_E < 101.2$~GeV. More recently, the ATLAS collaboration has performed a search for heavy charged leptons decaying to a $Z$ boson and an electron or muon, excluding the mass range 129–176~GeV (114–168~GeV) for mixing with only electrons (muons), except for the interval 144–163~GeV (153–160~GeV)~\cite{ATLAS:2015qoy}. In~\cite{Kumar:2015tna}, the prospects of probing singlet vector-like leptons with multi-lepton searches at the LHC were explored, with some reach for exclusion at the HL-LHC. In our specific model, the pair-produced vector-like leptons can also decay to a RH neutrino and the singly-charged scalar, $E^\pm\to N_i \phi^\pm$. The subsequent decay $\phi^{\pm}\rightarrow \ell^{\pm}\nu$ therefore leads to the signature of two oppositely charged leptons plus missing energy. This is a signature similar to that from $\phi^+\phi^-$ pair production and is expected to yield a similar constraint, $m_E \gtrsim 200$~GeV. 

Finally, EW precision data from LEP, complemented by the measurement of the Higgs mass, have been used to perform a global fit for the SM plus heavy NP effects, parametrised by $d = 6$ SMEFT operators~\cite{delAguila:2008pw, deBlas:2013gla,deBlas:2014mba}. From this fit, flavour-dependent upper bounds can be placed on the mixing between the charged leptons and $E$, which alternatively can be written as the lower limits $m_E/|Y_E^e| > 8.3$~TeV, $m_E/|Y_E^\mu| > 5.8$~TeV and $m_E/|Y_E^\tau| > 5.3$~TeV (95\%~C.L.). The singly-charged scalar is similarly constrained as $m_\phi/|f_{e\mu}|> 12.5$~TeV (95\%~C.L.).

\subsection{Constraints from Charged Lepton Flavour Violation}
\label{subsec:cLFV}

%
\begin{figure}[t!]
\centering
\includegraphics[width=0.24\textwidth]{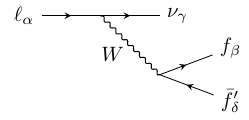}
\includegraphics[width=0.24\textwidth]{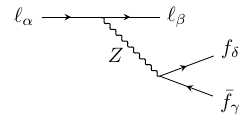}
\includegraphics[width=0.24\textwidth]{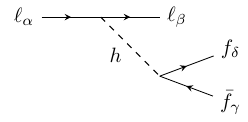}
\includegraphics[width=0.24\textwidth]{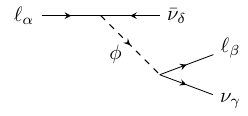}
\caption{Diagrams at tree-level in the UV extension which result in violations of LFU and non-vanishing rates for the cLFV processes $\ell_\alpha\to \ell_\beta \ell\bar{\ell}$ and $\tau\to \ell_\beta X$, where $X$ is a light meson.}
\label{fig:mu-to-3e}
\end{figure}

As mentioned in Section~\ref{sec:model}, charged lepton flavour violating (cLFV), lepton flavour universality (LFU) violating observables and precision tests of observables such as $(g-2)_\mu$ can be used to constrain the parameter space of the UV model~\cite{Falkowski:2013jya,Crivellin:2017rmk,Poh:2017tfo,Crivellin:2020klg,Crivellin:2020ebi,CarcamoHernandez:2021yev,Felkl:2021qdn,Hamaguchi:2022byw,Altmannshofer:2022fvz}. In the presence of the Yukawa couplings $Y_E$, mixing is induced between the SM charged leptons and the vector-like lepton $E$, modifying the SM charged-current, neutral-current and Higgs interactions at low energies, as seen in Eqs.~\eqref{eq:off-diag-Z-couplings} and \eqref{eq:off-diag_Higgs}.\footnote{The Yukawa coupling $Y_\nu$, which induces the active-sterile mixing $V_{\ell N}$, also modifies the charged- and neutral-current interactions and therefore contributes to cLFV processes~\cite{Alonso:2012ji}. For simplicity, we assume that $Y_{\nu} \ll 1$ for the purposes of deriving bounds on the UV model.}

Equivalently, from the EFT perspective, the operators $\mathcal{O}_{Hl}^{(1)}$, $\mathcal{O}_{Hl}^{(3)}$ and $\mathcal{O}_{eH}$ are induced at tree-level after integrating out $E$. Likewise, the operator $\mathcal{O}_{ll}$ is generated after integrating out $\phi$. SM processes such as $\ell_{\alpha}\to\ell_\beta\nu\bar{\nu}$, $\pi\to \ell_\alpha \nu$, $\tau\to \pi \nu$ and $B\to D^{(*)}\ell_\alpha\nu$ are modified in the presence of these operators, while cLFV processes such as $\ell_\alpha\to \ell_\beta\gamma$, $\ell_\alpha \to \ell_\beta\ell\bar{\ell}$, $\mu - e$ conversion in nuclei, $Z\to \ell_\alpha^{+}\ell_\beta^{-}$ and $h\to \ell_\alpha^{+}\ell_\beta^{-}$ are induced. The $\ell_\alpha\to \ell_\beta\gamma$ process additionally receives contributions directly from the dipole operators $\mathcal{O}_{eB}$ and $\mathcal{O}_{eW}$, both of which are generated after integrating out $E$ and $\phi$ at one-loop. In the following, we assume that $E$ and $\phi$ are integrated out at a scale well above the low energies of the considered processes. Thus, one should consider the running of the operators from the high scale down to the EW scale, match to $SU(3)_c \times U(1)_Y$ invariant operators, and then run down to the low scale. However, we neglect the sub-leading effects of running as we only aim to obtain bounds on the model for comparison with those from LLP searches in Section~\ref{sec:LLP}.

For the bounds from LFU violation, we consider here as an example only the purely leptonic LFU ratios, which measure deviations from the SM predictions for $\ell_\alpha\to \ell_\beta\nu\bar{\nu}$. In the UV model, we obtain the ratio~\cite{Crivellin:2020ebi,Allwicher:2021ndi}
\begin{align}
\label{eq:LFU_ratio_1}
\frac{\Gamma(\ell_\alpha \to \ell_\beta \nu\bar{\nu})}{\Gamma(\ell_\alpha \to \ell_\beta \nu\bar{\nu})\big|_{\text{SM}}} = 1 + v^2 \,\bigg[\frac{|Y_E^{\alpha}|^2 + |Y_E^{\beta}|^2}{2m_E^2} + \frac{2|f_{\alpha\beta}|^2}{m_\phi^2}\bigg]\,,
\end{align}
where we only take into account the interference between the SM and the NP contributions. The first and second terms in the square brackets of Eq.~\eqref{eq:LFU_ratio_1} arise from the exchange of $W$ and $\phi$, respectively, as seen in the left- and right-most diagrams of Fig.~\ref{fig:mu-to-3e}. The $Z$ and $h$ exchange diagrams do not contribute at this level, because there are no FCNCs at tree-level in the SM. This ratio can be used to determine the ratio of couplings $|g_\alpha/g_\beta|$, obtained as
\begin{align}
\label{eq:LFU_ratio_2}
\left|\frac{g_\alpha}{g_\beta}\right| \approx 1 + v^2 \,\bigg[\frac{|Y_E^{\alpha}|^2 - |Y_E^{\beta}|^2}{4m_E^2} + \frac{|f_{\alpha\rho}|^2 - |f_{\beta\rho}|^2}{m_\phi^2}\bigg]\,.
\end{align}
From Eq.~\eqref{eq:LFU_ratio_2}, it is clear that LFU violation vanishes for flavour universal couplings, e.g., the choice $Y_E^e = Y_E^\mu = Y_E^\tau$ and $f_{e\mu} = f_{e\tau} = f_{\mu\tau}$. Current experimental values of these coupling ratios are $|g_\tau/g_\mu| = 1.0009(14)$, $|g_\tau/g_e| = 1.0027(14)$, $|g_\mu/g_e| = 1.0019(14)$~\cite{Pich:2013lsa,HFLAV:2022esi}. From the first of these results, taking only $Y_E^\tau$ to be non-zero gives $m_E/|Y_E^{\tau}| > 4.1~\text{TeV}$.

The tree-level $Z$ and Higgs exchange diagrams in Fig.~\ref{fig:mu-to-3e} induce the cLFV processes $\ell_\alpha \to \ell_\beta\ell\bar{\ell}$, which are subject to stringent constraints from SINDRUM~\cite{SINDRUM:1987nra} and Belle~\cite{Hayasaka:2010np}. The branching ratio for this general process, neglecting final-state lepton masses, is~\cite{Kuno:1999jp,Crivellin:2013hpa}  
\begin{align}
\label{eq:Br_lto3l}
\mathcal{B}(\ell_\alpha\to \ell_\beta\ell_\gamma\bar{\ell}_\delta) &\approx \frac{m_\alpha^5}{768\pi^3\Gamma_\alpha(1+\delta_{\beta\gamma})}\frac{|Y_E^{\alpha *}Y_{E}^\beta|^2}{m_E^4}\left[(1+\delta_{\beta\gamma})(g_L^\ell)^2 + (g_{R}^\ell)^2\right]\,.
\end{align}
We assume that the dominant contribution to Eq.~\eqref{eq:Br_lto3l} comes from the $Z$ exchange diagrams, or correspondingly the operators $\mathcal{O}_{Hl}^{(1)}$ and $\mathcal{O}_{Hl}^{(3)}$. The contribution of the Higgs exchange diagram, or the operator $\mathcal{O}_{eH}$, can be safely neglected as it is further suppressed by the lepton masses, as seen in Eqs.~\eqref{eq:L_Higgs} and \eqref{eq:off-diag_Higgs}. In Eq.~\eqref{eq:Br_lto3l}, we also neglect the contributions from $E$ and $\phi$ at one-loop via penguin and box diagrams, or equivalently via the operators $\mathcal{O}_{eB}$, $\mathcal{O}_{eW}$ and $\mathcal{O}_{ll}$~\cite{Crivellin:2020klg}. The one-loop contribution of $\phi$ will only be relevant if the Yukawa couplings inducing $\ell_\alpha \to \ell_\beta\ell\bar{\ell}$ at tree-level satisfy $Y_E^\alpha, Y_E^\beta \ll f_{\alpha\beta}$. If we assume that the tree-level contribution dominates, the branching ratio in Eq.~\eqref{eq:Br_lto3l} and the corresponding experimental upper limits can be used to place lower bounds on the combination $m_E/|Y_{E}^{\alpha*}Y_{E}^{\beta}|^{1/2}$ for $\alpha\neq\beta$. The SINDRUM experiment provides the upper limit $\mathcal{B}(\mu^{+}\to e^+ e^+ e^-) < 1 \times 10^{-12}$ (90\%~C.L.), giving $m_E/|Y_{E}^{\mu*}Y_{E}^{e}|^{1/2} > 120~\text{TeV}$. Likewise, the upper limits $\mathcal{B}(\tau^- \to e^-e^-e^+) < 2.7\times 10^{-8}$ and $\mathcal{B}(\tau^- \to \mu^-\mu^-\mu^+) < 2.1\times 10^{-8}$ from Belle translate to $m_E/|Y_{E}^{\tau*}Y_{E}^{e}|^{1/2} > 5.9~\text{TeV}$ and $m_E/|Y_{E}^{\tau*}Y_{E}^{\mu}|^{1/2} > 6.3~\text{TeV}$, respectively. The equivalent constraints from $\mathcal{B}(\tau^{-} \to e^-\mu^-\mu^+)$ and $\mathcal{B}(\tau^- \to 
\mu^-e^-e^+)$ are comparable. We note that the cLFV processes $\ell_\alpha \to \ell_\beta q \bar{q}$, i.e., hadronic $\tau$ decays $\tau\to\ell_\beta X$, where $X$ is a light pseudoscalar, scalar or vector meson, are also generated. Constraints on such decay modes from Belle and BaBar are comparible to those on $\tau\to \ell_\beta \ell\bar{\ell}$~\cite{Banerjee:2022xuw}.

In the presence of the flavour-changing $Z$ couplings at tree-level, the exotic process of muon conversion to electrons in nuclei can also occur. The rate for this process, divided by the total capture rate $\Gamma_{\text{capt}}$, is~\cite{Kitano:2002mt,Davidson:2018kud}
\begin{align}
\label{eq:Cr_mutoe}
\text{CR}(\mu\to e) & = \frac{m_\mu^5}{\Gamma_{\text{capt}}}\frac{\big|Y_{E}^{\mu*}Y_{E}^{e}\big|^2}{m_E^4}\bigg|(g_L^u + g_R^u)(2V^{(p)}+V^{(n)})+(g_L^d + g_R^d)(V^{(p)}+2V^{(n)})\bigg|^2\,,
\end{align}
where $V^{(p)}$ and $V^{(n)}$ are nucleus-dependent overlap integrals over the proton and neutron densities and the muon and electron wavefunctions~\cite{Kitano:2002mt}. In Eq.~\eqref{eq:Cr_mutoe}, we again assume that the dominant contribution arises at tree-level via the operators $\mathcal{O}_{Hl}^{(1)}$ and $\mathcal{O}_{Hl}^{(3)}$, neglecting the contributions of $E$ and $\phi$ at one-loop via penguin diagrams. The SINDRUM experiment has placed an upper bound on the $\mu-e$ conversion rate on $^{197}_{~79}$Au of $\text{CR}(\mu\to e)< 7\times 10^{-13}$~(90\%~C.L.)~\cite{SINDRUMII:2006dvw}, which results in the constraint $m_E/|Y_{E}^{\mu*}Y_{E}^{e}|^{1/2} > 290~\text{TeV}$.

\begin{figure}[t!]
\centering
\includegraphics[width=0.24\textwidth]{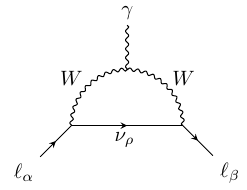}
\includegraphics[width=0.24\textwidth]{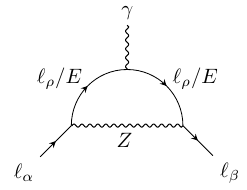}
\includegraphics[width=0.24\textwidth]{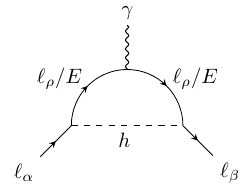}
\includegraphics[width=0.24\textwidth]{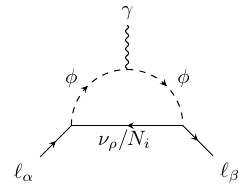}
\caption{Diagrams at one-loop in the UV model which induce the cLFV process $\ell_\alpha \to \ell_\beta \gamma$ and contribute to the anomalous magnetic moment of charged leptons, $a_\ell$. From left to right, the first three diagrams are the result of the modified $W$, $Z$ and Higgs interactions in Eqs.~\eqref{eq:L_gauge} and \eqref{eq:L_Higgs}, respectively, while the fourth diagram is induced by the $-\bar{L}f\tilde{L}\phi$ and $-\bar{N}^c_R f' \ell_R\phi^*$ interactions. The first and last diagrams are further modified in the presence of the active-sterile mixing $V_{\ell N}$.}
\label{fig:mu-to-e-gamma}
\end{figure}

Next, the radiative cLFV process $\ell_\alpha \to \ell_\beta \gamma$ is generated at one-loop, as depicted in Fig.~\ref{fig:mu-to-e-gamma}. The corresponding branching ratio is given 
by~\cite{Crivellin:2013hpa,Pruna:2014asa,Davidson:2016edt}
\begin{align}
\label{eq:Br_ltolgamma}
\mathcal{B}(\ell_\alpha\to \ell_\beta\gamma) & \approx \frac{m_\alpha^5}{512\pi^3\Gamma_\alpha}\frac{\alpha}{2\pi} \Bigg[\bigg|\frac{Y_E^{\alpha *}Y_E^{\beta}}{4m_E^2}\bigg(1 + \frac{8}{3} g_L^\ell\bigg) + \frac{f_{\rho\alpha}^*f_{\rho\beta}}{3m_\phi^2}\bigg|^2 + \bigg|\frac{f_{i\alpha}'f_{i\beta}^{\prime *}}{12 m_\phi^2}\bigg|^2\Bigg]\,,
\end{align}
where the indices $\rho$ and $i$ are summed over. The term in Eq.~\eqref{eq:Br_ltolgamma} proportional to $Y_E^{\alpha*}Y_E^\beta$ originates from the operators $\mathcal{O}_{Hl}^{(1)}$, $\mathcal{O}_{Hl}^{(3)}$, $\mathcal{O}_{eB}$ and $\mathcal{O}_{eW}$, generated after integrating out $E$, while the terms proportional to $f_{\rho\alpha}^* f_{\rho\beta}$ and $f_{i\alpha}' f_{i\beta}^{\prime *}$ arise from the contributions of $\phi$ to $\mathcal{O}_{eB}$ and $\mathcal{O}_{eW}$. Contributions of the operator $\mathcal{O}_{eH}$ via two-loop Barr-Zee type diagrams are neglected, as $C_{eH}$ is suppressed by the lepton masses. If we assume only non-zero $Y_E^\alpha$ values, the current upper limit from the MEG experiment~\cite{MEG:2016leq}, $\mathcal{B}(\mu\to e\gamma) < 4.2 \times 10^{-13}$ (90\%~C.L.), corresponds to the constraint $m_E/|Y_{E}^{\mu*}Y_{E}^{e}|^{1/2} > 75~\text{TeV}$. For $\tau$ decays, the BaBar experiment gives the upper limits $\mathcal{B}(\tau\to e\gamma) < 3.3 \times 10^{-8}$ and $\mathcal{B}(\tau\to \mu\gamma) < 4.4 \times 10^{-8}$ (90\%~C.L.)~\cite{BaBar:2009hkt}, enforcing $m_E/|Y_{E}^{\tau*}Y_{E}^{e}|^{1/2} > 2.9~\text{TeV}$ and $m_E/|Y_{E}^{\tau*}Y_{E}^{\mu}|^{1/2} > 2.7~\text{TeV}$, respectively.

Finally, limits can be placed from the cLFV decays $Z\to \ell_\alpha^{+}\ell_\beta^{-}$ and $h\to \ell_\alpha^{+}\ell_\beta^{-}$. With non-zero values of $Y_E^\alpha$, the branching ratios of these are~\cite{Harnik:2012pb,Blankenburg:2012ex,Falkowski:2013jya,Altmannshofer:2022fvz}
\begin{align}
\mathcal{B}(Z\to \ell_{\alpha}^+\ell_{\beta}^-) &= \frac{m_Z}{6\pi\Gamma_Z}\frac{m_Z^2}{v^2}\left[\left(\big(g_L^\ell\big)^2 + \big(g_R^\ell\big)^2 + g_L^\ell \frac{v^2|Y_{E}^\alpha|^2}{2m_E^2}\right)\delta_{\alpha\beta} + \frac{v^4 |Y_{E}^{\alpha *} Y_E^{\beta}|^2}{16m_E^4}\right]\,, \\
\mathcal{B}(h\to \ell_{\alpha}^+\ell_{\beta}^-) &= \frac{m_h}{8\pi\Gamma_h}\frac{m_\alpha^2}{v^2}\left[\left(1 - \frac{3 v^2|Y_{E}^\alpha|^2}{2m_E^2}\right)\delta_{\alpha\beta} + \frac{9 v^4 |Y_{E}^{\alpha *} Y_E^{\beta}|^2}{16m_E^4}\right]\,.
\end{align}
We note that $\mathcal{B}(Z\to \ell_{\alpha}^{\pm}\ell_{\beta}^\mp)$ and $\mathcal{B}(h\to \ell_{\alpha}^{\pm}\ell_{\beta}^\mp)$ are found by adding the branching ratios with $\alpha \leftrightarrow\beta$ to those above. Upper bounds on these decays have been placed by OPAL~\cite{OPAL:1995grn}, ATLAS~\cite{ATLAS:2014vur,ATLAS:2019pmk} and CMS~\cite{CMS:2021rsq}. Taking the most stringent, $\mathcal{B}(Z\to e^\pm \mu^\mp) < 4.2\times 10^{-7}$ at 90\%~C.L.~\cite{ATLAS:2014vur}, gives $m_E/|Y_{E}^{\mu*}Y_{E}^{e}|^{1/2} > 4.1~\text{TeV}$.

To conclude this section, we observe that $\ell_\alpha \to \ell_\beta\ell\bar{\ell}$ and $\mu-e$ conversion in nuclei provide the most stringent constraints on the UV model, with the latter probing values of $m_E$ up to 290~TeV for $\mathcal{O}(1)$ Yukawa couplings $Y_E$. The coupling $Y_E^\alpha$ enters the active-to-sterile dipole couplings, as seen in Eq.~\eqref{eq:CNB_matching} and \eqref{eq:CNW_matching}, and therefore these constraints must be taken into account when considering non-zero $d_{\s \nu N\gamma}^{\alpha i}$. However, we note that these constraints can be easily evaded if the couplings of $E$ and $\phi$ are not flavour universal. For example, one can consider only non-zero couplings in the $\mu$-$\tau$ ($e$-$\tau$) sector; for $Y_E^e = 0$ ($Y_{E}^\mu = 0$), the strong bounds from $\mu\to 3e$ and $\mu - e$ conversion are no longer applicable. The only bounds that then apply are the much weaker limits from LFU violation, $\tau\to \mu\gamma$ ($\tau\to e\gamma$), and $\tau\to 3\mu$ ($\tau\to 3e$). One can also consider the case where only one Yukawa coupling is non-zero, e.g. $Y_E^\tau$. Then the bounds strongest bounds come from the one-loop contributions of $\phi$ to $\mu\to e\gamma$, $\mu\to 3e$ and $\mu- e$ conversion.

\section{Long-Lived Particle Searches at the LHC using Non-Pointing Photons}
{\label{sec:LLP}}

In this section, we discuss the potential of searches for non-pointing photons at the LHC to constrain sterile neutrino magnetic moments. To keep our analysis as general as possible, here we will follow the same conventions as the EFT
notation introduced in Section~\ref{sec:EFT}.

\subsection{Sterile Neutrino Production and Decay Mechanisms}

In what follows, we first discuss the sterile neutrino production and decay processes triggered by the EFT Wilson coefficients $C_{\s{NNB}}^{(5)ij}$ and $C_{\s{NX}}^{(6)\alpha i}$, where $X=B,W$.\footnote{Here, we assume $C_{\s{NW}}^{(6)\alpha i } = 2/(3 t_w) C_{\s{NB}}^{(6)\alpha i}$, as discussed at the end of Section~\ref{sec:EFT}.} We also comment on the contributions from the active-sterile neutrino mixing.

Hereafter we simplify notation by omitting flavour indices. First, we take $C_{\s{NNB}}^{(5)12} \equiv C_{\s{NNB}}^{(5)}$ and since $\mathcal{O}_{\s{NNB}}^{(5)}$ is an antisymmetric operator, we are implicitly assuming $C_{\s{NNB}}^{(5)21} = - C_{\s{NNB}}^{(5)12}$. Second, we also omit the lepton flavour $\alpha$ in $C_{\s{NX}}^{(6)\alpha i}$, $(C_{\s{NX}}^{(6)\alpha i} \equiv C_{\s{N_i X}}^{(6)})$, as processes which involve active neutrinos do not reveal which neutrino flavour is participating. These simplifications will also apply to the coefficients in the broken phase.

\begin{figure}[t]
\centering
    \includegraphics[width=0.48\textwidth]{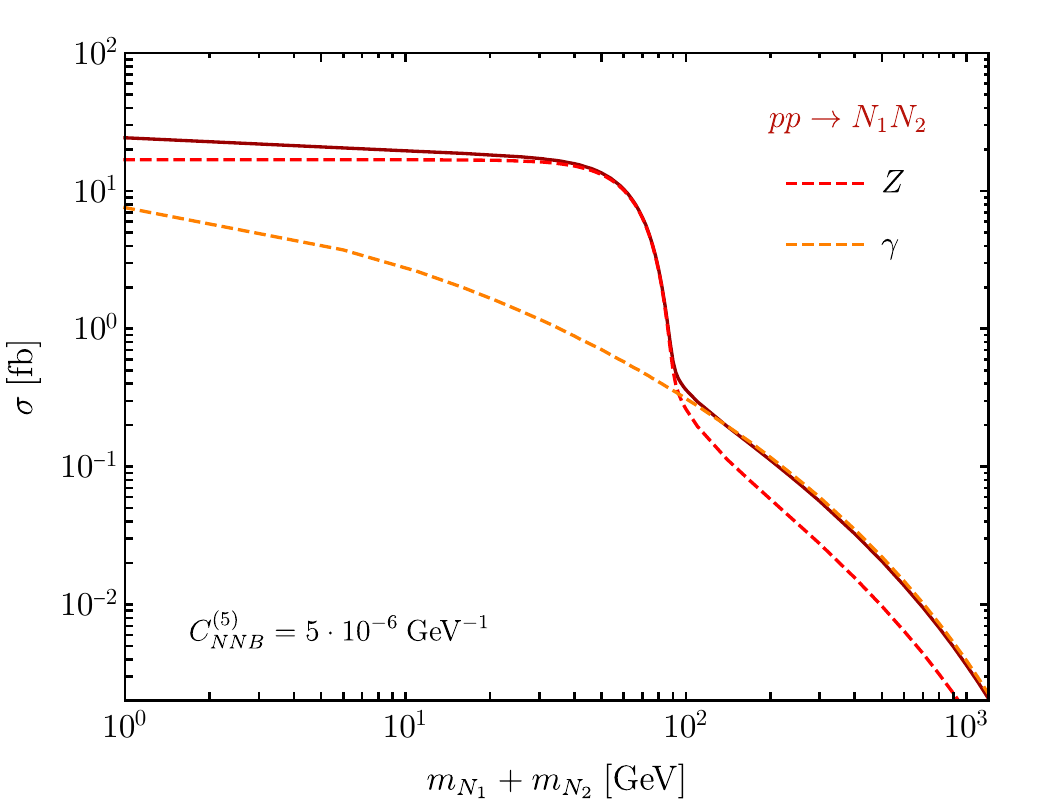}
    \includegraphics[width=0.48\textwidth]{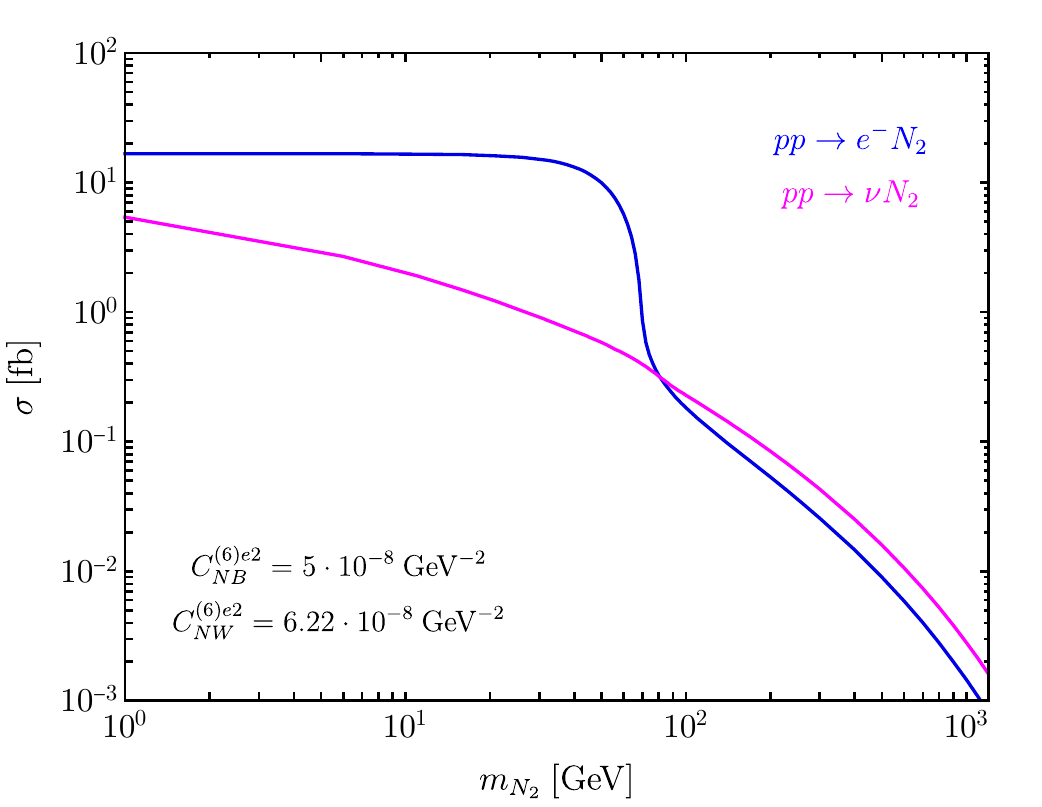}
    \caption{Example cross-sections at the LHC ($\sqrt{s}=14$~TeV). The left plot shows the contribution of $C_{\s NNB}^{(5)}$ to $\sigma(pp\to N_1 N_2)$ as a function of $m_{N_1}+m_{N_2}$. Dashed lines represent the partial contributions from the $Z$ and $\gamma$ channels. The right plot displays the production cross-section of the processes mediated by $C_{\s{NB}}^{(6)}$ and $C_{\s{NW}}^{(6)}$. The cross-section for the charged-current process given by the chosen values of these couplings is the same as the cross-section for the production of $N_2$ via the mixing $|V_{e N_2}|^2 \simeq 3\times 10^{-6}$.}
\label{fig:xsection}
\end{figure}

The coupling $C_{\s{NNB}}^{(5)}$ contributes to pair-production of the sterile neutrinos via $pp\rightarrow \gamma/Z \rightarrow N_1 N_2$, while $C_{\s{N_iX}}^{(6)}$ leads to the production of a single $N_i$ through $pp\rightarrow \gamma/Z \rightarrow N_i \nu$ and $pp \rightarrow W^\pm \rightarrow N_i \ell^\pm$. Fig.~\ref{fig:xsection} shows some example cross-sections for fixed values of $C_{\s{NNB}}^{(5)}$ (left) and $C_{\s{N_iX}}^{(6)}$ (right).
For $m_{N_1}+m_{N_2} < m_Z $, pair production through $C_{\s{NNB}}^{(5)}$ is dominated by the $Z$ exchange diagram, whereas for larger masses photon exchange dominates. Single sterile neutrino production through $C_{\s{N_iX}}^{(6)}$, on the other hand, is dominated
by the charged current. The neutral channel is subdominant compared to the charged one in this case because of the suppression between $d_{\s \nu N_i \gamma}$ and $d_{\s \nu N_i Z}$ discussed below Eq.~\eqref{eq:dnuNratio}. Note that sterile neutrino production at the LHC through mixing (non-zero $V_{\ell N_i}$) proceeds through charged
current events with the same dependence on the mass $m_{N_i}$ as for $C_{\s{N_iX}}^{(6)}$. We therefore do not show production via mixing separately in Fig.~\ref{fig:xsection} (right).

For sterile neutrino decays, the interaction $C_{\s{NNB}}^{(5)}$ induces the two-body decay $N_2 \rightarrow N_1 \gamma$ and, if $N_2$ is heavy enough, $ N_ 2 \rightarrow N_1 Z$. Note that the lightest $N_1$ can not decay through $C_{\s{NNB}}^{(5)}$.  The relevant partial decay widths of $N_2$, in terms of the broken phase parameters defined in \eqref{OP:ZpoleET}, are given by
\begin{equation}
\label{eq:DecN2N1g}
\Gamma(N_2 \rightarrow N_1 \gamma) = \frac{2 |d_{\s NN\gamma}|^2}{\pi} m_{N_2}^3
\bigg( 1 - \frac{m_{N_1}^2}{m_{N_2}^2} \bigg)^3 = \frac{2 |d_{\s NN\gamma}|^2}{\pi} m_{N_2}^3
\left( 2 - \delta \right)^3 \delta^3\,,
\end{equation}
with $\delta\equiv 1- \frac{m_{N_1}}{m_{N_2}}$ denoting the mass splitting between the two sterile states,
and
\begin{equation}
\label{eq:DecN2N1Z}
\Gamma(N_2 \rightarrow N_1 Z) = \frac{ |d_{\s NNZ}|^2}{\pi}\, m_{N_2}^3\, f_Z
\left(\frac{m_{N_1}}{m_{N_2}}, \frac{m_Z}{m_{N_2}}\right) \, \lambda^{1/2}
\bigg(1, \frac{m_{N_1}^2}{m_{N_2}^2},\frac{m_Z^2}{m_{N_2}^2}\bigg)\,,
\end{equation}
with 
\begin{eqnarray}
f_Z(x,y) &=& \left((1-x)^2-y^2\right)\left(2(1+ x)^2 +y^2 \right)\,, \\
\lambda(x,y,z) & =& (x-y-z)^2 - 4yz\,.
\end{eqnarray}
We recall that we use the notation $d_{\s NN\gamma}^{12} \equiv d_{\s NN\gamma}$ in this section, and because of the antisymmetric nature of the coupling, there is an additional factor of 4 in the decay width of $N_2 \to N_1 \gamma$ compared to that of $N_i \to \nu \gamma$ (see Eqs.~\eqref{eq:DecN2N1g} and \eqref{eq:DecN1nugam}).

The couplings $C_{\s{N_iB}}^{(6)}$ and $C_{\s{N_iW}}^{(6)}$ generate the decays $N_i \rightarrow \nu \gamma$, $N_i \rightarrow \nu Z$ and $N_i \rightarrow \ell^\pm W^\mp$. While the $\gamma$ final state is kinematically allowed for practically all values of $m_{N_i}$, $Z$ and $W$ final states contribute only for $m_{N_i}$ larger than the gauge boson masses.\footnote{For $m_{N_i}$ below the weak gauge boson masses, three-body decays through off-shell $Z/W$ occur, but we have checked that these
are always subdominant in comparison to the $\gamma$ channel. Hence these partial widths are neglected in our numerical study.} The relevant partial decay widths, written in terms of the broken phase parameters defined in Eqs.~\eqref{OP:ZpoleET_b} and \eqref{OP:ZpoleET_W}, are given by
\begin{eqnarray}\label{eq:DecN1nugam}
\Gamma(N_i \rightarrow \nu \gamma) &=& \frac{|d_{\s \nu N_i \gamma}|^2}{2\pi}
 m_{N_i}^3,\\ \label{eq:DecN1nuZ}
\Gamma(N_i \rightarrow \nu Z) &=& \frac{|d_{\s \nu N_i Z}|^2}{8 \pi } m_{N_i}
(2m_{N_i}^2 + m_Z^2) \lambda\bigg(1,0,\frac{m_Z^2}{m_{N_i}^2}\bigg),\\
\Gamma(N_i \rightarrow \ell^- W^+) &=& \frac{|d_{\s \ell N_i W}|^2}{16\pi} 
m_{N_i} (2m_{N_i}^2 + m_W^2) \lambda\bigg(1,0,\frac{m_W^2}{m_{N_i}^2}\bigg),
\end{eqnarray}
where, for simplicity, we have neglected the charged lepton masses. In the computation of the total width of a Majorana $N_i$, the latter two partial decay widths are multiplied by 2 to account for the Majorana nature of $N$ and the charged conjugated channel $\ell^+ W^-$. Sterile neutrinos will also decay via mixing. These decays have been thoroughly studied in the literature. For our numerical analysis, we use the decay width formulas of~\cite{Atre:2009rg, Bondarenko:2018ptm}.
  
\begin{figure}[t]
\centering
    \includegraphics[width=0.49\textwidth]{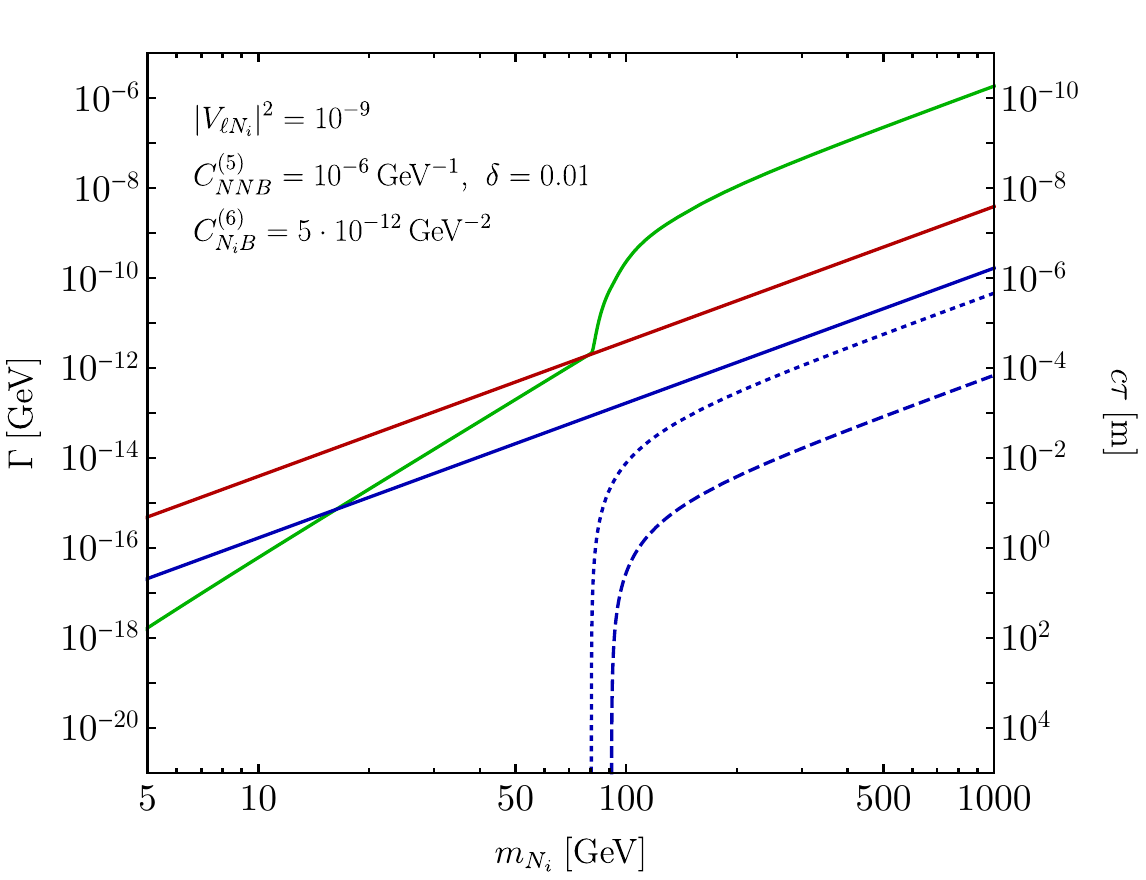}
    \includegraphics[width=0.49\textwidth]{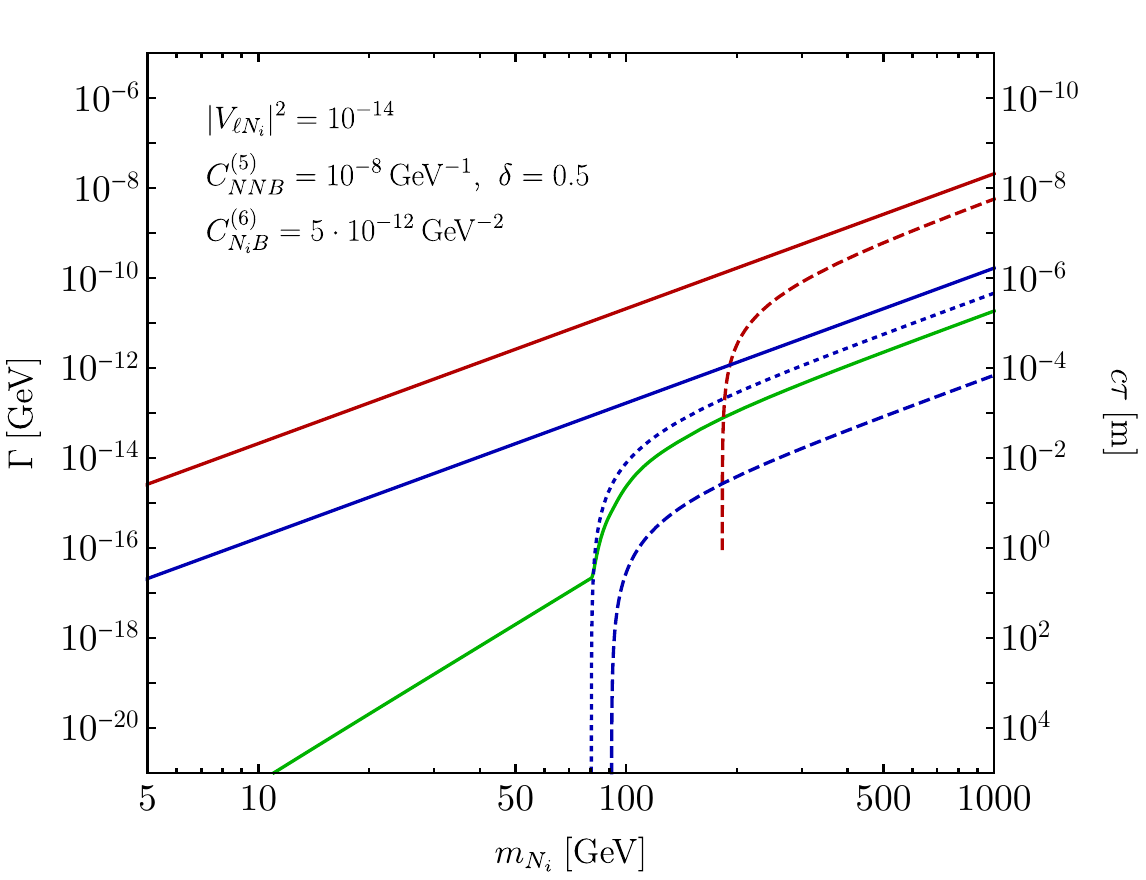}
    \includegraphics[width=0.4\textwidth]{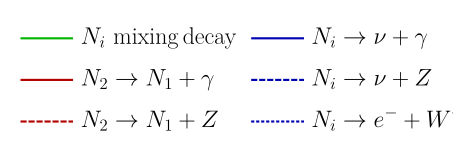}
\caption{Partial decay widths of the decay modes induced by neutrino mixing, $C_{\s{NNB}}^{(5)}$ and $C_{\s{N_iB}}^{(6)}$ ($C_{\s{N_iW}}^{(6)} = 2/(3 t_w) C_{\s{N_iB}}^{(6)}$). We have chosen different benchmark values of the coefficients in each plot, see the corresponding legend. }
\label{fig:decaywidth}
\end{figure}

In Fig.~\ref{fig:decaywidth} we show the previous partial decay widths as a function of the sterile neutrino masses for various coefficient values.
Here, $C_{\s{N_iB}}^{(6)}$ and $C_{\s{N_iW}}^{(6)}$  are fixed as $C_{\s{N_iB}}^{(6)} = 5 \times 10^{-12}$~GeV$^{-2}$ and $C_{\s{N_iW}}^{(6)} = 2/(3t_w)  C_{\s{N_iB}}^{(6)} \approx 6.22 \times 10^{-12}$~GeV$^{-2}$. These control the decay channels to $\nu \gamma$, $\nu Z$, $e^- W^+$. 
For comparison, we also show the neutrino mixing partial decay width for two different values of $|V_{\ell N_i}|^2$. In the left (right) plot, we fix $|V_{\ell N_i}|^2 = 10^{-9}$ $(10^{-14})$, motivated by an inverse seesaw (or naive Type-I seesaw) estimation. 
Additionally, we show the partial decay width of $N_2 \to N_1 \gamma$ and $N_2 \to N_1 Z$ controlled by $C_{\s{NNB}}^{(5)}$ and $\delta$.
In the left plot, these parameters have been fixed to $C_{\s{NNB}}^{(5)} = 10^{-6}$~GeV$^{-1}$ and $\delta = 0.01$, and in the right plot, to $C_{\s{NNB}}^{(5)} = 10^{-8}$~GeV$^{-1}$ and $\delta = 0.5$. These values are chosen to illustrate the regimes in which sterile neutrinos ranging from a few GeV to a few hundred GeV could have decay lengths $\mathcal{O}(10^{-2}-1)$~m, which are interesting for LLP searches.
Notice that only the right plot shows a curve for $N_2 \to N_1 Z$, as a large mass splitting is needed for this channel.

\begin{figure}[t]
\centering
\includegraphics[width=0.95\textwidth]{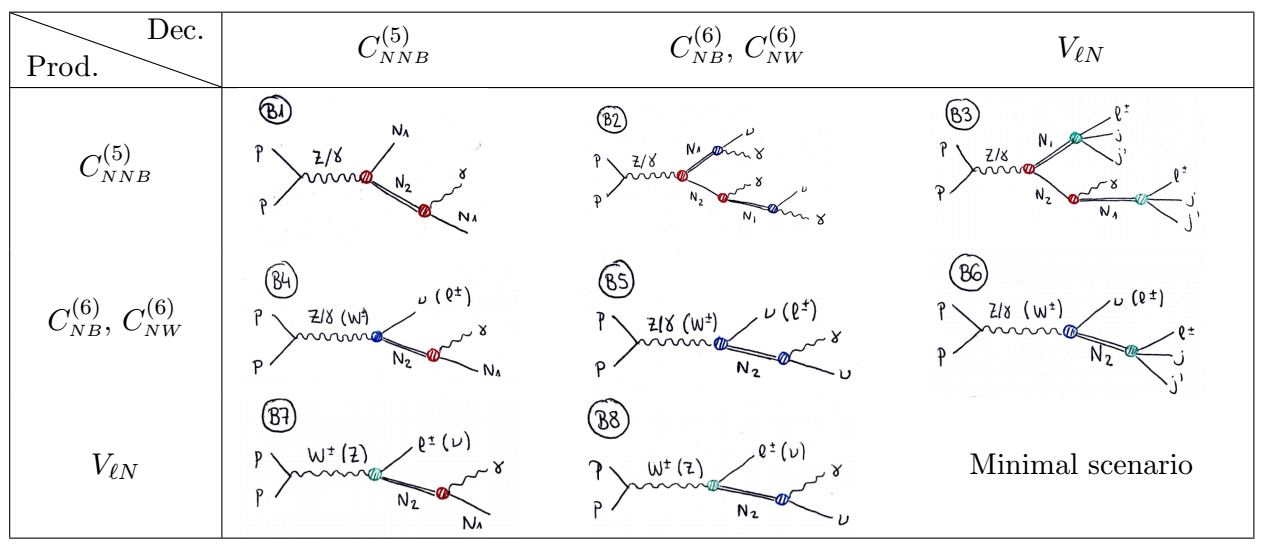}
\caption{Nine different process classes: This classification distinguishes scenarios, depending on which couplings dominate $N$ production and decay.  The sterile-to-sterile (active-to-sterile) magnetic moment is controlled by $C_{\s{NNB}}^{(5)}$ ($C_{\s{NB}}^{(6)}$, $C_{\s{NW}}^{(6)}$). The active-sterile neutrino mixing is parameterised by $V_{\ell N}$.  
  }
\label{tab:Benchmarks}
\end{figure}

From Figs.~\ref{fig:xsection} and \ref{fig:decaywidth} we can compare the relative sizes of $C_{\s{NNB}}^{(5)}$, $C_{\s{N_iX}}^{(6)}$ and $V_{\ell N_i}$, determining the dominant sterile neutrino production and decay mechanisms. There are nine distinct possibilities, which we define as shown in Fig.~\ref{tab:Benchmarks}. The scenarios in the same row have the same dominant production mechanism, whereas the scenarios in the same column share the same dominant decay mode for the sterile neutrino. In practice, not all combinations will lead to non-pointing photons and we present the features of all these scenarios below.

Scenarios \textbf{B1}-\textbf{B3} assume that production is controlled by $C_{\s{NNB}}^{(5)}$, thus $N_2$ and $N_1$ are pair
produced. Then:
\begin{itemize}
\item \textbf{B1}: $N_2$ will decay to $N_1\gamma$, whereas $N_1$ escapes undetected. The decay length of $N_2$ is controlled by $C_{\s{NNB}}^{(5)}$ and the mass difference $\delta = 1 - m_{N_1} /m_{N_2}$. $N_2$ is long-lived if $\delta \ll 1$.  Too small $\delta$, however, leads to photons too soft to be tagged. The signal consists of one non-pointing photon plus missing energy.    \item \textbf{B2}: $N_1$ will decay to $\nu \gamma$ and for small enough $C_{\s{N_1X}}^{(6)}$ this decay is long-lived. $N_2$ will decay to $N_1\gamma$ either promptly or, if $\delta \ll 1$, with a finite decay length. The signature of this scenario contains up to three photons, either two or three of which are non-pointing. There is also missing energy in the event, from $\nu$ and the potential undetected photons.
\item \textbf{B3}: The mixing parameter $V_{\ell N_1}$ controls the decay of $N_1$. Again, $N_2$ will decay to $N_1\gamma$ promptly (or delayed if $\delta \ll 1$). The potential final state signal is a prompt (or non-pointing) photon plus one or two displaced vertices (DV) with charged tracks from the $N_1$ decays. Unless two displaced vertices and the photon are all found, the event again contains missing energy.
\end{itemize}
In scenarios \textbf{B4}-\textbf{B6}, production is
controlled by $C_{\s{N_iX}}^{(6)}$: 
\begin{itemize}      
\item \textbf{B4}: Comparing the production cross-sections in Fig.~\ref{fig:xsection} to the decay widths in Eqs.~\eqref{eq:DecN2N1g} and \eqref{eq:DecN1nugam} it becomes quickly clear that this scenario can not be realised, since $C_{\s{NNB}}^{(5)}$ large enough to dominate the decay width over $C_{\s{N_iX}}^{(6)}$ would also dominate the cross-section.
\item \textbf{B5}: In this scenario $C_{\s{N_iX}}^{(6)}$ is assumed to be also dominant in the decay of $N_i$. Comparing the width \eqref{eq:DecN1nugam} to the production cross-section in Fig.~\ref{fig:xsection} (right), it is clear that the decays $N_i\to \nu \gamma$ are too fast to lead to non-pointing photons since there is no kinematic suppression ($\delta \simeq 1$) if $C_{\s{N_iX}}^{(6)}$ is large enough to give a sizeable cross-section. The signal in this case is a prompt lepton accompanied by a prompt photon.
\item \textbf{B6}: $V_{\ell N_i}$ is assumed to dominate the decay. Given the discussion above for (B5), it is clear that decays will be prompt in this case. There are no photons in the final state, thus no difference in signal to the minimal mixing case.
\end{itemize}
Finally, for \textbf{B7}-\textbf{B9} we assume production is dominated by mixing, $V_{\ell N_i}$. 

\begin{itemize}
 \item \textbf{B7}: The conditions defining this scenario can be
  fulfilled only in a narrow range of parameters. The cross-section
  from the dipole and the one from mixing are roughly of the same
  order for some particular ratio of $C_{\s{NNB}}^{(5)}$ to $V_{\ell N_i}$.
  For this ratio, the width from $C_{\s{NNB}}^{(5)}$ will be smaller
  than the one from mixing for $m_{N_2}>15$ GeV, for
  $\delta=0.01$. For a smaller ratio (and thus a more dominant
  production via mixing), the partial width through the dipole can
  dominate only for smaller values of $m_{N_2}$. In this narrow
  parameter range some events with a non-pointing photon would be generated, however, always together with a certain number of
  displaced vertex events.
\item \textbf{B8}: The decay caused by $C_{\s{N_iX}}^{(6)}$ can easily dominate over the decay via mixing for $m_{N_i}<m_Z$, without $C_{\s{N_iX}}^{(6)}$ dominating the cross-section. Since a minimum size of $V_{\ell N_i}$ is
  needed to give a sizeable cross-section, however, decay lengths will be shorter than in the pure mixing scenario. The signal is a prompt lepton plus a photon (prompt or non-pointing, depending
  on $C_{\s{N_iX}}^{(6)}$).
\item \textbf{B9}: This scenario is phenomenologically the same as the minimal mixing scenario. It has been discussed extensively in the literature.
\end{itemize}

Given the above discussion, the most interesting scenarios are
\textbf{B1}-\textbf{B3}. We will therefore study these three scenarios in detail below. 

\subsection{Simulation Details}{\label{sec:simulation}}
In order to determine the sensitivity reach of ATLAS for probing sterile neutrino magnetic moments, we performed a numerical study of the representative chosen benchmarks, using Monte-Carlo (MC) techniques. We first implemented our model described by Eq.~\eqref{Lag:SMEFT$+N_R$} (only up to $d = 6$) plus the interactions induced by active-sterile neutrino mixing in \verb|FeynRules| \cite{Christensen:2008py,Alloul:2013bka}. The generated UFO files \cite{Degrande:2011ua} were embedded in \verb|MadGraph5| \cite{Alwall:2014hca}, where we generated the collision process $p p \rightarrow N_1 N_2$ at $\sqrt{s} = 14$\,TeV, induced by $C_{\s{NNB}}^{(5)}$ in the selected benchmarks. We generated $10^5$ events at each point of the grid covering the parameter plane $(m_{\text{LLP}}, c_{\text{decay}})$, where the LLP is the sterile neutrino providing the displaced signature at the detector and $c_{\text{decay}}$ denotes the interaction dominating the LLP decay width. These two parameters define the two-dimensional space we scan over in our simulation for each benchmark scenario, as shown in the second column of Tab.~\ref{tab:simulation}. In the third column, we display the remaining parameters of the model, which are fixed to some numerical value. In B1, we keep the mass splitting $\delta$ fixed such that $m_{N_1}$ varies in the scan along with $m_{N_2}$. Whereas in B2 and B3 we fix $m_{N_2}$ and $C_{\s{NNB}}^{(5)}$, since they do not affect the long-lived nature of $N_1$.

\begin{table}[]
\centering
\renewcommand{\arraystretch}{1.3}
\begin{tabular}{|c|cc|c|}
\hline
\multirow{2}{*}{Scenario} & \multicolumn{2}{c|}{Model parameters} & \multirow{2}{*}{Simulated decay} \\ \cline{2-3}
 & \multicolumn{1}{c|}{Scan} & Fixed &  \\ \hline \hline
B1 & \multicolumn{1}{c|}{$m_{N_2}$, $C_{\s{NNB}}^{(5)}$} & $\delta$ & $N_2 \rightarrow N_1 \gamma$ \\ \hline
B2 & \multicolumn{1}{c|}{$m_{N_1}$, $C_{\s{N_1X}}^{(6)}$} & $m_{N_2}$, $C_{\s{NNB}}^{(5)}$ & \begin{tabular}[c]{@{}c@{}}$N_2 \rightarrow N_1 \gamma$ \\ $N_{1} \rightarrow \nu \gamma$\end{tabular} \\ \hline
B3 & \multicolumn{1}{c|}{$m_{N_1}$, $|V_{eN_1}|^2$} & $m_{N_2}$, $C_{\s{NNB}}^{(5)}$ & \begin{tabular}[c]{@{}c@{}}$N_2 \rightarrow N_1 \gamma$ \\ $N_1 \rightarrow e j j $\end{tabular} \\ \hline
\end{tabular}
\caption{Parameters entering the numerical simulation in each benchmark. The scan parameters define the two-dimensional space to be surveyed, while the other model parameters remain fixed. The last column indicates the decay channel studied in the simulation. }
\label{tab:simulation}
\end{table}

The sterile neutrino decays are treated separately in \verb|MadSpin| \cite{Artoisenet:2012st}. To ensure numerical stability for small decay widths, the $N_i$ was forced to decay into the final state of interest (shown in the last column of Tab.~\ref{tab:simulation}) in all simulated events. The decay LHE event files, given as an output by \verb|MadSpin|, were then processed by \verb|Pythia8| \cite{Sjostrand:2014zea} to estimate the ATLAS efficiency in detecting the characteristic LLP signature of each benchmark: non-pointing photons in B1 and B2, and displaced vertices in B3. Specific details of each search strategy are outlined below.

Non-pointing photons are emitted in the decay of the long-lived sterile neutrinos. These decays occur at a secondary vertex, displaced from the collision point or primary vertex (PV), causing the photon direction to point away from the PV. The ATLAS (and CMS) Electromagnetic Calorimeter (ECal) can detect energetic photons and reconstruct their trajectories precisely. The photon displacement is quantified using the impact parameter (IP), defined as the minimum distance of the photon trajectory to the PV. The IP can be decomposed into its transverse and longitudinal components, $d_{\s{XY}}$ and $d_{\s{Z}}$, which can be obtained with
\begin{align}
    & d_{\s{XY}} = x_{\s{LLP}} \frac{p_{\s{Y}}}{p_{\s{T}}} -   y_{\s{LLP}} \frac{p_{\s{X}}}{p_{\s{T}}}, \\[0.5em]
    & d_{\s{Z}} =  \frac{z_{\s{LLP}}-(\vec{r}\cdot\vec{p}) p_{\s{Z}} /|\vec{p}|^2}{1- p_{\s{Z}}^2/|\vec{p}|^2}  ,
\end{align}
where $\vec{r}=\{x_{\s{LLP}}, y_{\s{LLP}}, z_{\s{LLP}}\}$ is the LLP decay position and $\vec{p}$ is the photon momentum. 

In our simulation, we have access to the true decay positions of the sterile neutrinos as well as to the kinematic properties of the outgoing particles. Therefore, we can use the previous equations to compute $d_{\s{XY}}$ or $d_{\s{Z}}$ for each photon emitted in the decay of $N_i$. In practice, measuring $d_{\s{Z}}$ requires knowing the exact location of the PV along the beamline ($z$-axis). However, this can be challenging in the high-luminosity conditions at the LHC, where multiple collisions occur per bunch crossing, leading to the production of several PVs. Moreover, in our specific case (scenarios B1-B3), sterile neutrinos are produced at the collision point without any accompanying charged particle. This limitation extends, in general, to scenarios where only neutral particles are produced at the origin and they decay far from the PV. Therefore, we can only use the variable $d_{\s{XY}}$, which can be measured with respect to the beamline. 

It is worth mentioning that recent works have been performed in this direction. ATLAS, for instance, presented results from a search using non-pointing and delayed photons in \cite{ATLAS:2022vhr}, and in~\cite{Duarte:2023tdw, Delgado:2022fea}, the authors have recast it for constraining Higgs decays into sterile neutrinos in the context of dimension-5 operators. However, the search relies on $d_Z$ and the time delay $t_\gamma$, which also requires knowing the PV location. We cannot apply this search to our selected benchmarks and accordingly propose an alternative search strategy. Our approach draws inspiration from the CMS search in~\cite{CMS:2012idp}, which looked for non-pointing photons using instead a trigger on $d_{\s{XY}}$.

In Tab.~\ref{tab:cuts}, we summarise the cuts for the non-pointing photon search that we apply in scenarios B1 and B2. The event selection starts by triggering photons satisfying $|p_T^\gamma|>10$~GeV and $|\eta^\gamma|<2.47$. Subsequently, we demand the LLP to decay before it reaches the outer layer of the ECal. This requirement imposes the following cuts on the transverse plane and longitudinal axis: $r_{\text{DV}}<1450$~mm and $|z_{\text{DV}}|<3450$~mm, respectively. Finally, the energetic photons passing the initial selection criteria and originating from one of the DVs are required to have $|d_{\s{XY}}^\gamma|> 6$~mm. This last cut is expected to reduce SM background significantly \cite{CMS:2012idp}. In Fig.~3 of this paper, the CMS data indicates that backgrounds decrease exponentially with the transverse impact parameter, with less than one background event remaining after a cut of $6-7$ mm. However, since the CMS search used only 2.2~fb$^{-1}$ of statistics, a larger cut will be necessary for the high-luminosity LHC. This adjustment will primarily affect the upper parts of the sensitivity curves (in the plane coupling vs. mass), as the upper limit is determined by the LLP being too short-lived to leave a measurably displaced photon. Strengthening the cut on  $|d_{\s{XY}}^\gamma|$ to larger minimum values would reduce the sensitivity curves from the top. Predicting the exact value of this cut is challenging due to the predominantly instrumental nature of the background, thus we will restrict ourselves to the original 6~mm cut.

On the other hand, in scenario B3, the LLP candidate decays via the active-sterile neutrino mixing. The dominant decay modes contain charged particles that leave tracks that originate from a displaced vertex (DV). ATLAS and CMS have looked for long-lived sterile neutrinos using a DV search strategy and have placed bounds on the neutrino mixing parameter \cite{ATLAS:2022atq, CMS:2022fut}. Here, we follow~\cite{Cottin:2021lzz} and use their DV search, which targets a heavy $N$ decaying into $ejj$. We summarise the main selection cuts of this search in the lower part of Tab.~\ref{tab:cuts}. The event selection starts by identifying electrons with $|p_T^e| > 120$~GeV, $|\eta^e|<2.47$. Then we select events with a DV lying inside the ATLAS inner tracker detector, which imposes a cut in the transverse and longitudinal positions: $4$~mm $< r_{\text{DV}}<300$~mm,  $|z_{\text{DV}}|<300$~mm. To reconstruct the DV, at least four displaced charged particle tracks are needed. We require them to have a transverse impact parameter of $|d_0|>2$~mm.\footnote{The approximate transverse impact parameter in this case is defined as $d_0=r \cdot \Delta \phi$, where $r$ is the transverse distance of the track from the interaction point, and $\Delta \phi$ is the azimuthal angle between the track and the direciton of $N_1$.} Additionally, one of the displaced tracks must correspond to the energetic electrons passing the initial trigger. A final cut on the invariant mass of the DV ($m_{\text{DV}}>5$~GeV) is applied to remove SM background from B-mesons. Further details can be found in Section~III of \cite{Cottin:2021lzz}. 

\begin{table}[]
\centering
\renewcommand*{\arraystretch}{1.3}
\begin{tabular}{|c|c|c|}
\hline
 Scenario & Signature & Selection cuts  \\ \hline \hline
B1                  & Non-pointing $\gamma$          & $|p_T^\gamma| > 10$~GeV ,  $|\eta^\gamma|<2.47$        \\ \cline{1-2}
\multirow{2}{*}{B2} &  Non-pointing $\gamma$ ($\times 2$) & \multicolumn{1}{c|}{$r_{\text{DV}}<1450$~mm , $|z_{\text{DV}}|<3450$~mm} \\
                    &  (+ prompt $\gamma$)                 & $|d_{\s{XY}}^\gamma|> 6$~mm                                              \\ \hline
\multirow{4}{*}{B3} &      & $|p_T^e| > 120$~GeV, $|\eta^e|<2.47$                                \\
                    &     Displaced Vertex ($\times 2$)              & $4$~mm $< r_{\text{DV}}<300$~mm,  $|z_{\text{DV}}|<300$~mm                \\
                    &   (+ prompt $\gamma$)                                 & 4 tracks with $|d_0|>2$~mm  \\
                    &  & $m_{\text{DV}}> 5$~GeV   \\ \hline
\end{tabular}
\caption{Summary of selection cuts for a non-pointing photon search (B1 and B2) and a displaced vertex search (B3) at the ATLAS detector. }
\label{tab:cuts}
\end{table}

The total number of signal events for the three selected benchmarks is obtained with the following expressions:
\begin{equation}
    N_{\text{sig.}}^{\text{B1}} = \sigma\cdot \mathcal{L} \cdot \mathcal{B}\left( N_2  \rightarrow  N_1 \gamma\right) \cdot \epsilon_{\text{sel}}^{\text{B1}},
\end{equation}
\begin{equation}
    N_{\text{sig.}}^{\text{B2}} = \sigma \cdot \mathcal{L} \cdot \mathcal{B}\left( N_2  \rightarrow  N_1 \gamma\right) \cdot 2 \cdot \mathcal{B}\left( N_1  \rightarrow  \nu \gamma\right) \cdot \epsilon_{\text{sel}}^{\text{B2}},
\end{equation}
\begin{equation}
    N_{\text{sig.}}^{\text{B3}} = \sigma \cdot \mathcal{L} \cdot \mathcal{B}\left( N_2  \rightarrow  N_1 \gamma\right) \cdot 2 \cdot \mathcal{B}\left( N_1  \rightarrow  e j j \right) \cdot \epsilon_{\text{sel}}^{\text{B3}},
\end{equation}
where $\sigma$ is the cross-section of the process $pp\rightarrow N_1 N_2$ that depends on the sterile neutrino masses and the interaction $C_{\s{NNB}}^{(5)}$. $\mathcal{L}$ corresponds to the total integrated luminosity during the high-luminosity phase, $3$~ab$^{-1}$. The branching ratio into the appropriate final state is also a function of the masses and the coupling responsible for the decay in each scenario (see second column in Tab.~\ref{tab:simulation}). The factor $\epsilon_{\text{sel}}$ denotes the efficiency of event selection in ATLAS after applying the cuts in Tab.~\ref{tab:cuts}. Notice that there is an additional factor of two in the formula for B2 and B3 since we only require one displaced photon/vertex and there are two $N_1$ in each event.

Under the assumption of zero background, we derive the $95$\% C.L. sensitivity prospects of ATLAS by requiring 3 signal events.\footnote{ As discussed previously, the zero background assumption might be too optimistic for the cut of $|d_{\s{XY}}^\gamma|>6$~mm at the high-luminosity LHC. For this reason, we will also show lines for 10 or 30 events in the sensitivity plots, corresponding to roughly 25 and 225 background events for $95$\% C.L. sensitivities.} Exclusion limits can be placed in the corresponding parameter space if no signal events are found at the end of the experiment. However, in case of a discovery, a larger number of events would be necessary to identify the scenario. For instance, looking for a second photon in the non-pointing photon search could distinguish a potential signal event from B1 and B2. In scenario B3, on top of the displaced vertex search, an additional trigger on a prompt photon associated with the event containing the DV would distinguish this scenario from other models, such as the minimal scenario. 
\section{Results and Discussion}{\label{sec:results}}
Following the methodology described above, we obtain the experimental sensitivity of ATLAS for probing sterile neutrinos interacting via dipole interactions. We work at the level of $N_R$SMEFT operators in the unbroken phase and then make the conversion to the broken phase parameters using Eqs.~\eqref{OP:ZpoleET} and \eqref{OP:ZpoleET_b}. 
Besides, we consider the active-sterile neutrino mixing as an independent parameter and, for simplicity, focus on the mixing of the light sterile neutrino with the electron sector only, i.e. $V_{\ell N_1} = V_{eN_1}$. The scenarios under study are B1-B3 in Fig.~\ref{tab:Benchmarks}.
In these scenarios, sterile neutrinos are pair-produced in proton-proton collisions through the operator $\mathcal{O}_{\s{NNB}}^{(5)}$. The difference stems from the interaction dominating the long-lived sterile neutrino decays: $C_{\s{NNB}}^{(5)}$,  $C_{\s{N_1X}}^{(6)}$ ($X=B, W$) or $V_{eN_1}$. 

\begin{figure}[t]
\centering
\includegraphics[width=0.6\textwidth]{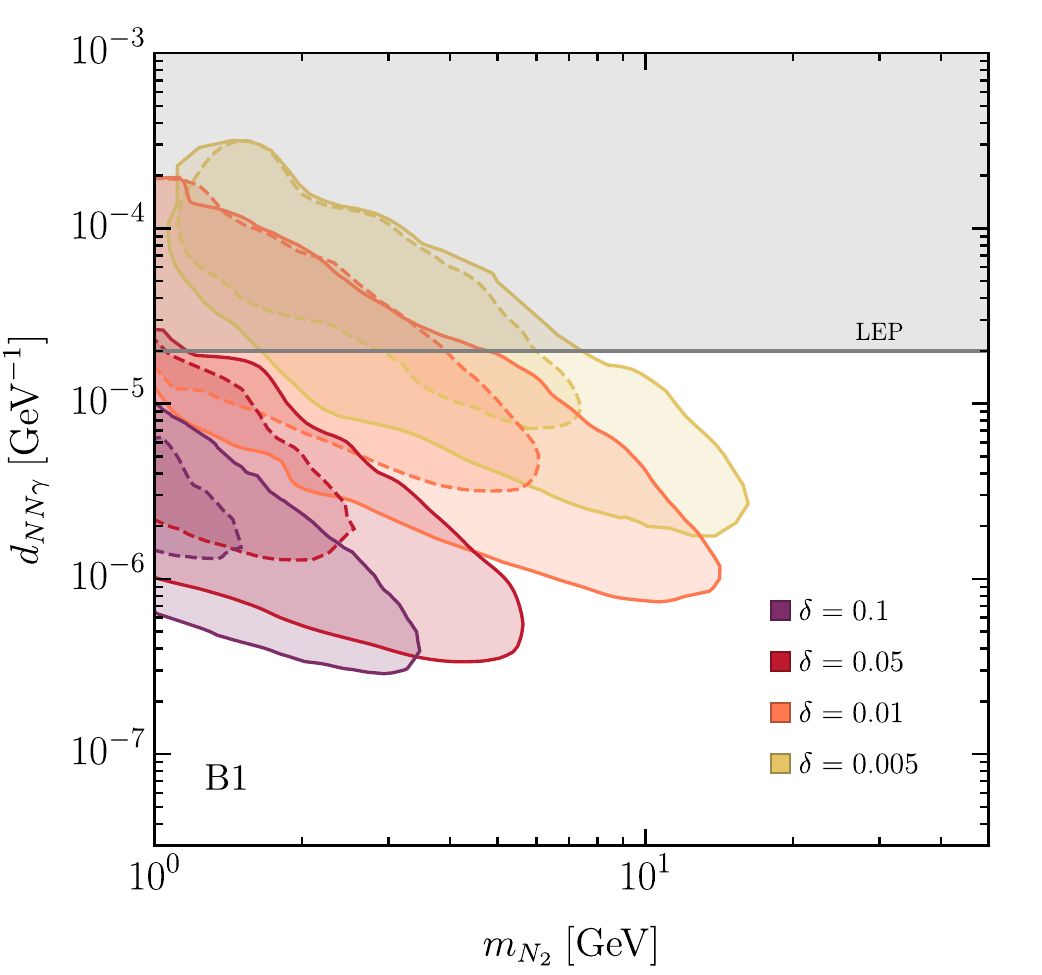} 
\caption{Sensitivity prospects of the non-pointing photon search at ATLAS ($\sqrt{s}=14$~TeV, 3~ab$^{-1}$) for scenario B1 $\left(pp\rightarrow N_1 N_2 \rightarrow N_1 (N_1 \gamma)^{\text{LLP}}\right)$. The 3 (30) event contours are shown as solid (dashed) lines in the plane $m_{N_2}$ vs. $d_{\s NN\gamma}$ for four values of the mass splitting $\delta= 1 - m_{N_1}/m_{N_2}$. The grey area shows the current constraint from LEP \cite{Magill:2018jla}.}
\label{fig:results_B1}
\end{figure}

Fig.~\ref{fig:results_B1} shows the sensitivity reach of the non-pointing photon search applied to scenario B1, in which $N_2$ is the LLP decaying within the detector. Contours corresponding to 3 (30) events are shown as solid (dashed) lines in the $m_{N_2}$~vs~$d_{\s NN\gamma}$ plane, for four mass splittings: $\delta=0.1$,  $\delta = 0.05$, $\delta = 0.01$ and $\delta = 0.005$. The sensitivity in mass goes beyond $m_{N_2} \simeq 3$~GeV, $m_{N_2} \simeq 5$~GeV, $m_{N_2}\simeq 14$~GeV and $m_{N_2}\simeq 16$~GeV, respectively. Dipole couplings as small as $d_{\s NN\gamma} \simeq 3 \times 10^{-7}$~$(8\times 10^{-7})$~GeV$^{-1}$ can be probed if $\delta=0.1$ ($0.01$), improving the current limits from LEP~\cite{Magill:2018jla} (shown as the grey area) by more than one order of magnitude in the corresponding mass ranges. For $\delta= 0.005$, the sensitivity reaches $d_{\s NN\gamma} \simeq 2 \times 10^{-6}$~GeV$^{-1}$. The shape of the curve can be understood in the following way. The lower part of the curve is the long-lived limit, for a given mass and $\delta$, a smaller coupling results in $N_2$ being too long-lived, escaping the detector without decaying. The upper part represents the opposite situation. A larger coupling value would make the $N_2$ decay too promptly and the emitted photon would point back to the primary vertex, losing the LLP signature. Moreover, since the production cross-section is proportional to $|d_{\s{NN\gamma}}|^2$, we quickly lose events as the dipole coefficient becomes smaller. However, we determined that the results are highly sensitive to the $p_{T}^\gamma$ cut in this scenario. The smaller the mass splitting, the less energetic the final photons are and for $\delta \lesssim 0.005$  they become too soft to be detected in the ECal. This effect can already be observed in the top left corner of the contour for $\delta=0.005$, where the search is no longer sensitive to low sterile neutrino masses.

\begin{figure}[t]
\centering
\includegraphics[width=0.6\textwidth]{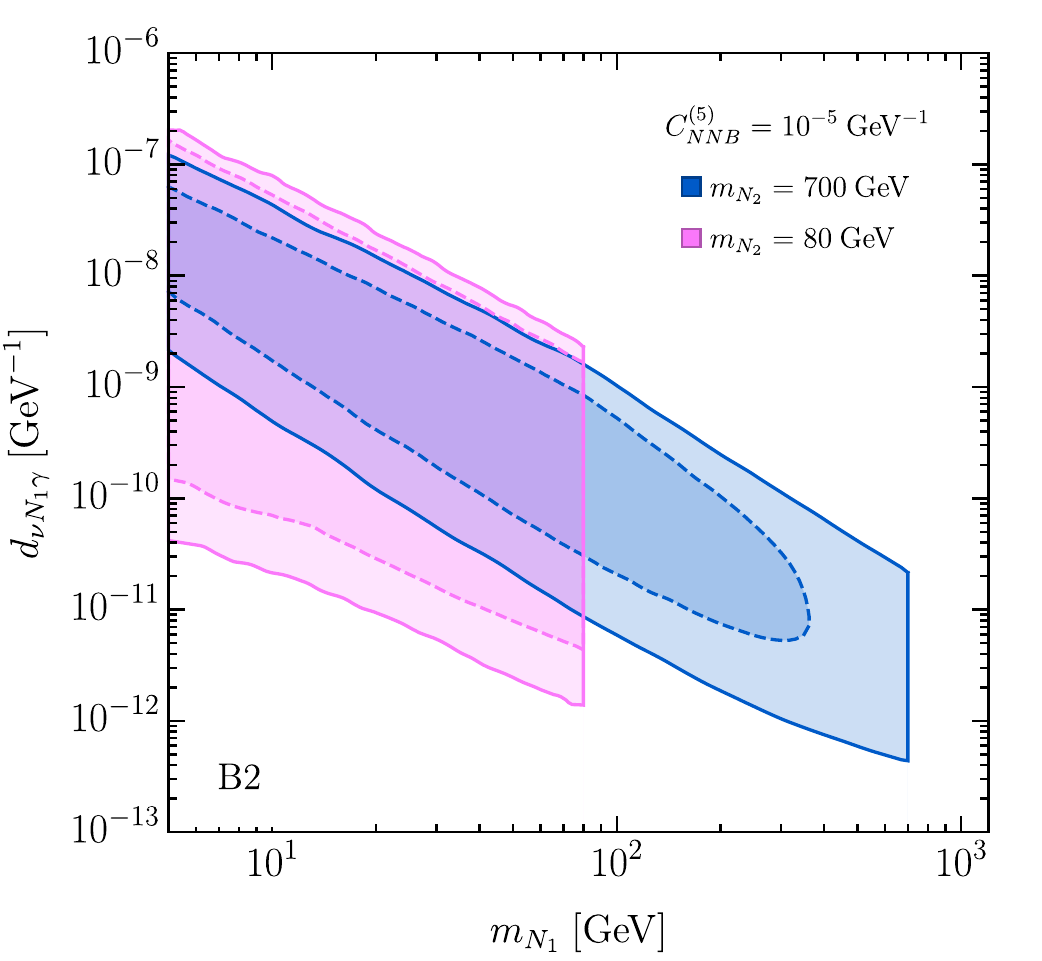} 
\caption{ Sensitivity prospects of the non-pointing photon search at ATLAS ($\sqrt{s}=14$~TeV, 3~ab$^{-1}$) for scenario B2: $pp \to N_1 N_2(\to N_1 \gamma) \to (\nu\gamma+ \nu\gamma)^{\text{LLP}} \gamma$. Solid (dashed) lines corresponding to 3 (30) events are shown in the $m_{N_1}$~vs.~$d_{\s \nu N \gamma}$ plane for fixed values of $m_{N_2}$ and $C_{\s{NNB}}^{(5)}$. The vertical lines in the upper mass reach come from the kinematic threshold ($m_{N_1}<m_{N_2}$).  Current exclusion bounds are weaker and are not shown.}
\label{fig:results_B2}
\end{figure}

Fig.~\ref{fig:results_B2} shows the results for scenario B2 in the $m_{N_1}$~vs.~$d_{\s \nu N_1 \gamma}$ plane. In this case, production and decay of the LLP candidate, $N_1$, are decoupled since they depend on different interactions. We fix $C_{\s{NNB}}^{(5)} = 10^{-5}$~GeV$^{-1}$, corresponding to $d_{\s{NN\gamma}}\simeq 8.8 \times 10^{-6}$~GeV$^{-1}$,
to ensure it dominates sterile neutrino production. Two mass values of $N_2$ are considered: $m_{N_2}=700$~GeV and $m_{N_2}=80$~GeV. We recall that the scan parameters are $m_{N_1}$ and $C_{\s{N_1B}}^{(6)}$, which then we convert onto $d_{\s \nu N_1 \gamma}$ using Eq.~\eqref{OP:ZpoleET_b}.\footnote{The contribution of $d_{\s \nu N_1 Z}$ and $d_{\s \ell N_1 W}$ to the total decay width of $N_1$ are also taken into account, since we are considering the EFT operators in the unbroken phase.} The solid (dashed) lines correspond to 3 (30) events. Equivalently, these dashed lines correspond to 3 events for $C_{\s NNB}^{(5)} \approx 3 \times 10^{-6}$~GeV$^{-1}$. The shape of the curves is similar to that of Fig.~\ref{fig:results_B1}, but the probed parameter space is much larger, precisely because the production and decay mechanisms of $N_1$ are decoupled. The sensitivity reach in mass is limited by the kinematic threshold $m_{N_1} \lesssim m_{N_2}$, resulting in these vertical lines at $m_{N_1}=700$~GeV and $m_{N_1} = 80$~GeV. As concerns active-to-sterile magnetic moments, values as small as $d_{\s \nu N_1 \gamma}\simeq 5 \times 10^{-13}~(2\times10^{-12})$~GeV$^{-1}$ can be probed for $m_{N_2}=700~(80)$~GeV. 
In contrast to the previous scenario, the non-pointing photons are considerably more energetic since they arise in the decay $N_1 \rightarrow \nu \gamma$. Consequently, almost all simulated photons pass the corresponding cuts of the non-pointing photon search, giving a large event selection efficiency. Even with a stronger $p_T^\gamma$ cut, the area enclosed by the sensitivity curves comprises unexplored parameter space. Indeed, current exclusion bounds are much weaker and do not appear in the figure. 

\begin{figure}[t]
\centering
\includegraphics[width=0.6\textwidth]{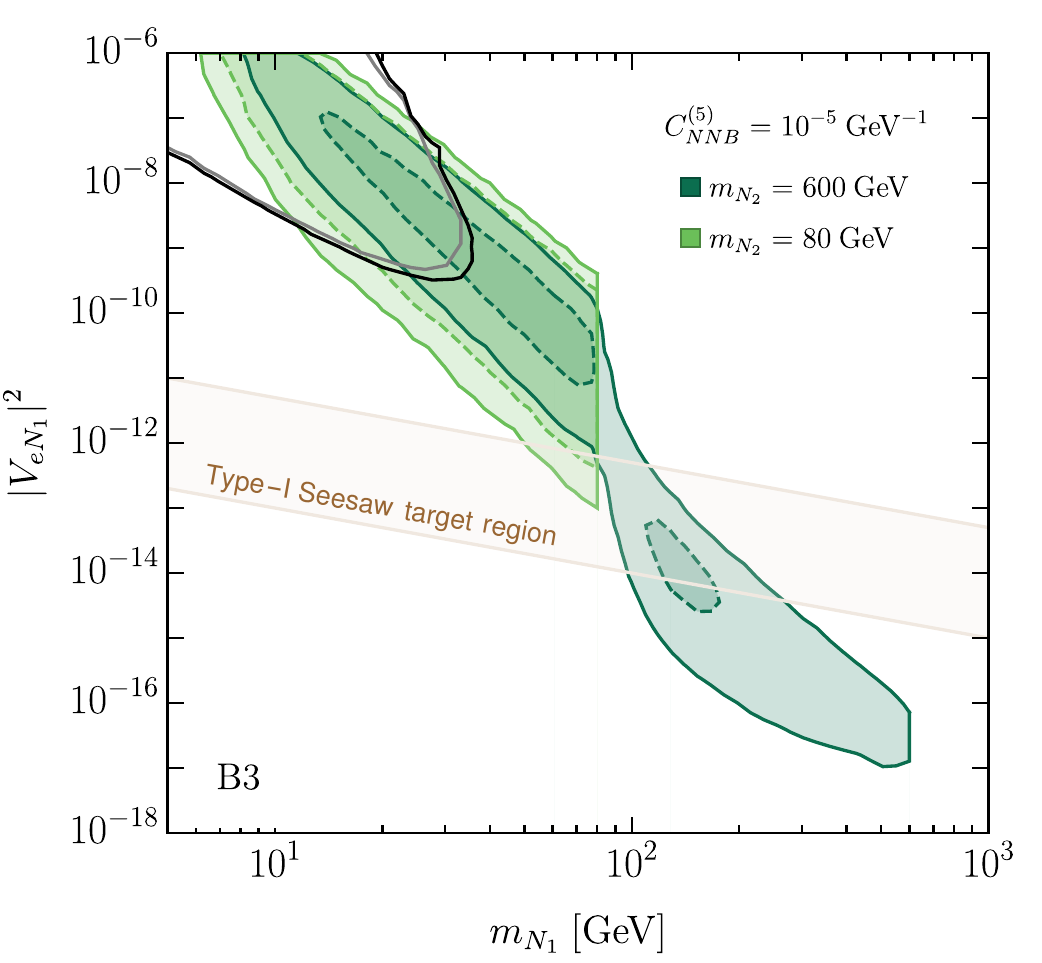} 
\caption{ Sensitivity prospects of the displaced vertex search at ATLAS ($\sqrt{s}=14$~TeV, 3~ab$^{-1}$) for scenario B3: $pp \to N_1 N_2(\to N_1 \gamma) \to (ejj+ejj)^{\text{LLP}} \gamma$. Solid (dashed) lines corresponding to 3 (10) events are shown in the $m_{N_1}$~vs.~$|V_{eN}|^2$ plane for fixed values of $m_{N_2}$ and $C_{\s{NNB}}^{(5)}$. The kinematic threshold ($m_{N_1}<m_{N_2}$) cuts the upper reach in mass with the vertical lines. The black and grey solid lines represent the CMS and ATLAS sensitivity prospects of a DV + prompt lepton search \cite{Drewes:2019fou, Cottin:2018nms,Beltran:2021hpq}. }
\label{fig:results_B3}
\end{figure}

For scenario B3, the results of the displaced vertex search are shown in Fig.~\ref{fig:results_B3}. We display the 3 (solid) and 10 (dashed) event contours in the $m_{N_1}$~vs.~$|V_{e N_1}|^2$ plane. Analogously to B2, production and decay of $N_1$ are decoupled. We fix the production coupling ($C_{\s{NNB}}^{(5)}=10^{-5}$~GeV$^{-1}$) and choose two values for $m_{N_2}$. For $m_{N_2} = 600$~GeV, the sensitivity extends across 10 orders of magnitude in $|V_{eN_1}|^2$, reaching down to $10^{-17}$, and saturates the kinematic threshold of $m_{N_1} \simeq 600$~GeV. The slope of the curve changes around $m_{N_1} \approx m_W$ because the $N_1$ can decay to on-shell weak bosons via the two-body decays $N_1\to W^\pm e^\mp$ and $N_1\to Z \nu$. This increase in the decay width is compensated by smaller values of the mixing parameter. For the chosen parameter values, 30 events are not reached and in the 10 event contour, we already observe two isolated regions in the parameter space.
Conversely, for $m_{N_2}=80$~GeV, 30 signal events can be obtained although we do not show the corresponding curve in the plot. The probed region, in this case, is one order of magnitude larger in $|V_{eN_1}|^2$ than for $m_{N_2}=600$~GeV, reaching values as small as $|V_{eN_1}|\simeq 10^{-13}$ at $m_{N_1}\simeq 80$~GeV.
Present exclusion bounds on the mixing parameter $|V_{eN}|^2$ are weaker in this sterile neutrino mass range and hence are not shown. However, we display the sensitivity prospects of displaced vertex searches by ATLAS and CMS \cite{Drewes:2019fou, Cottin:2018nms, Beltran:2021hpq}, represented with the grey and black lines. Their results are derived in the context of the minimal scenario (only neutrino mixing) and, to be precise, these searches cannot be directly recast into our parameter space as they require prompt leptons at the origin. Nevertheless, we show them for illustrative purposes, revealing that much smaller squared mixing values can be explored in the presence of sizeable sterile neutrino dipole moments, even accessing the Type-I seesaw band. This light grey region in the plot is obtained assuming the naive Type-I seesaw relation for active neutrino masses of $0.05$ and $0.001$~eV.

\begin{figure}[t]
\centering
\includegraphics[width=0.49\textwidth]{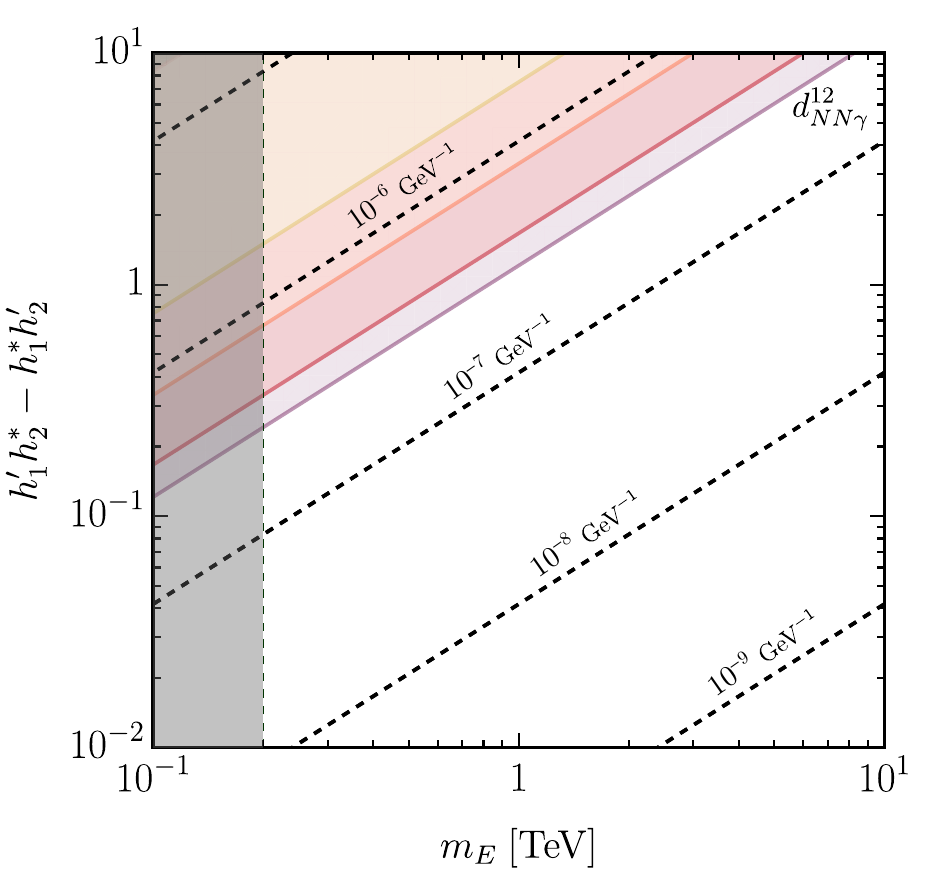} 
\includegraphics[width=0.49\textwidth]{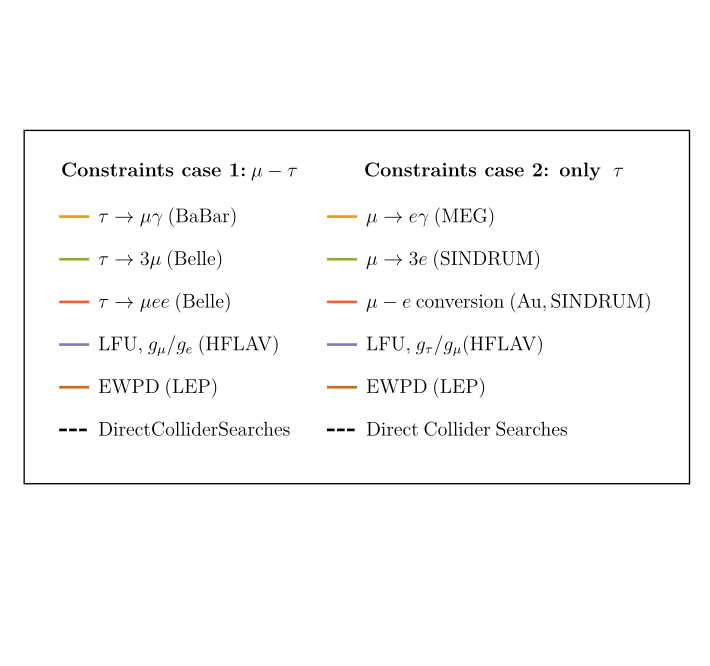} 

\includegraphics[width=0.49\textwidth]{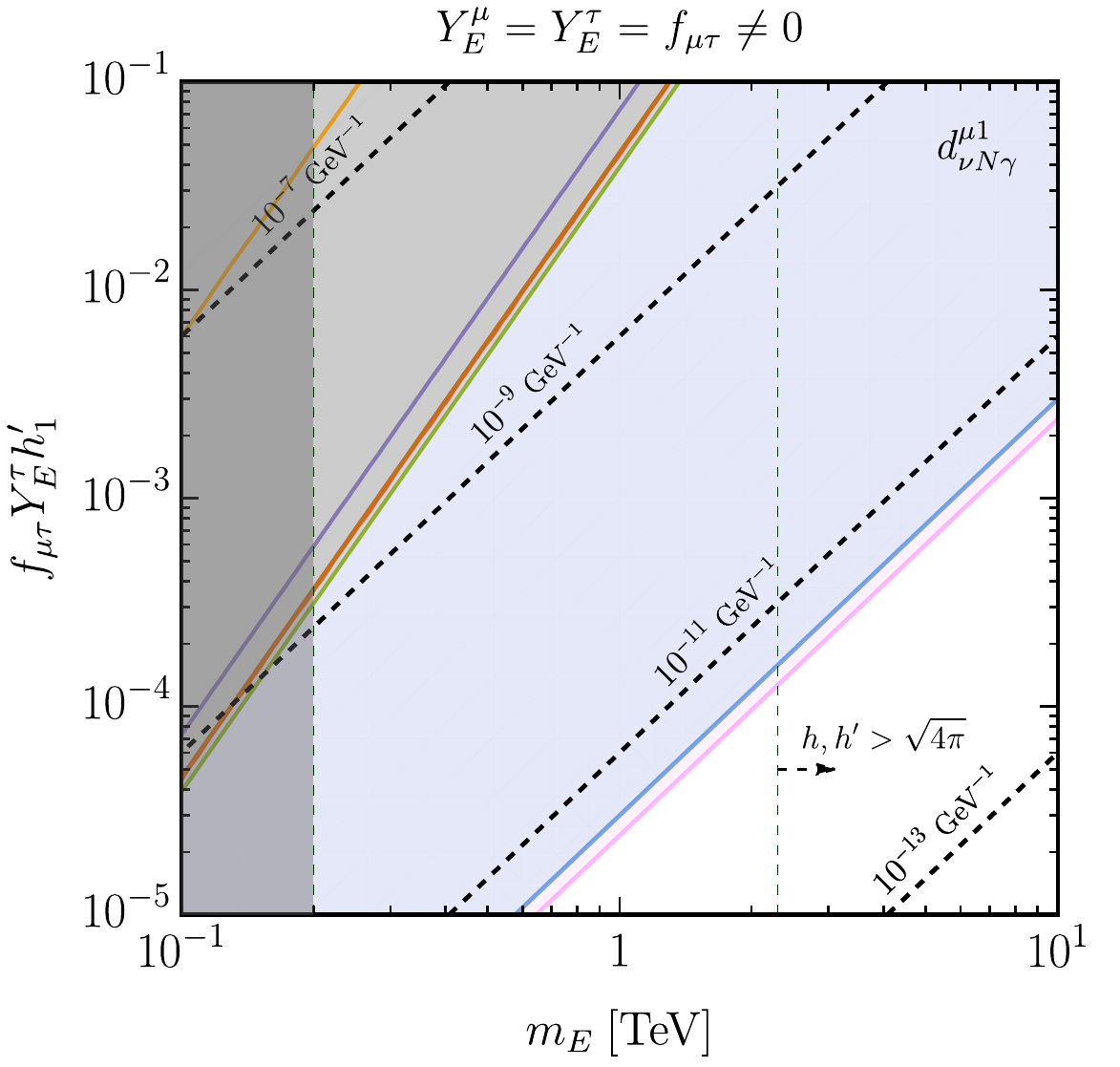} 
\includegraphics[width=0.49\textwidth]{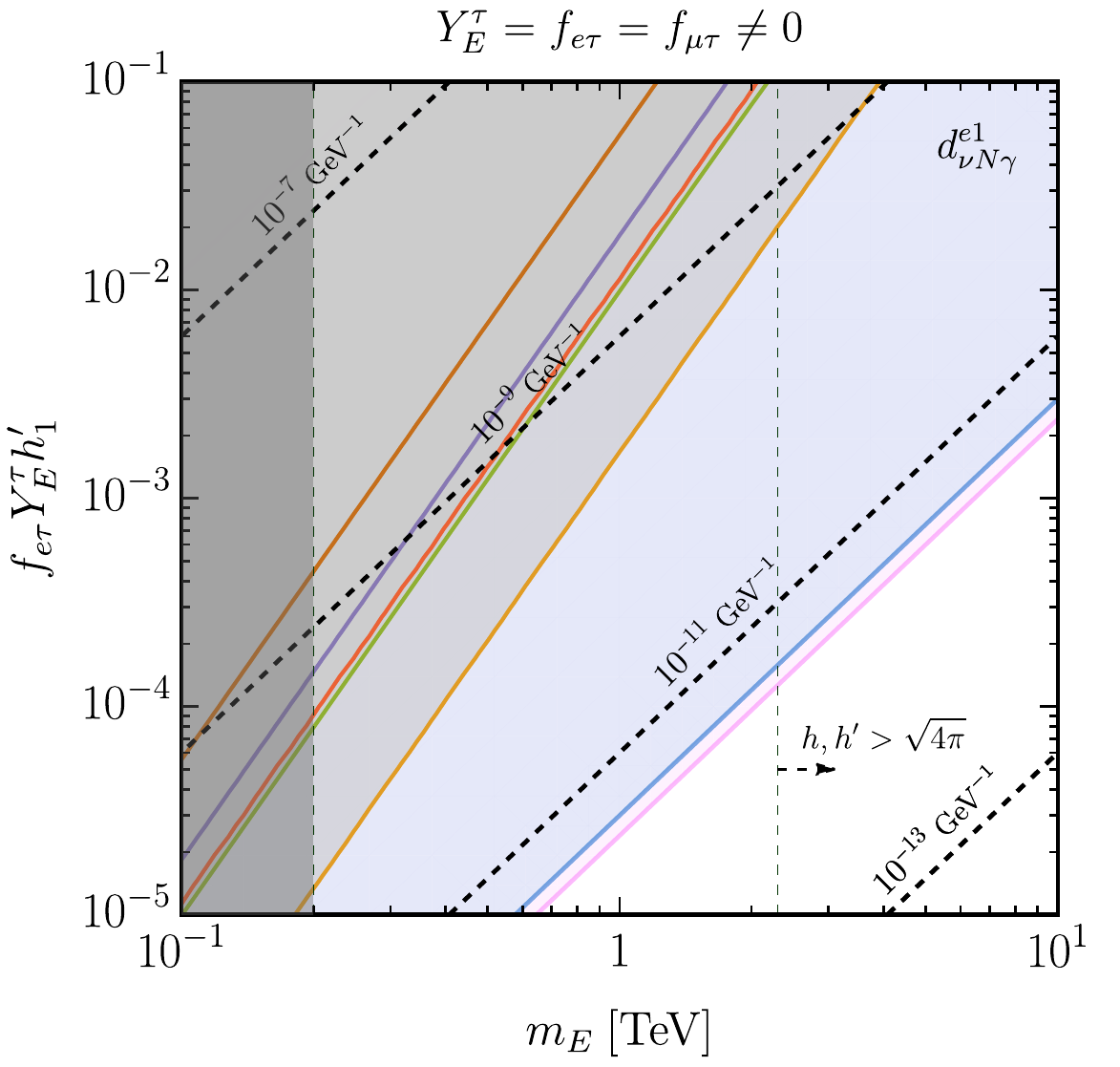} 
\caption{\textbf{Top left panel}: Sterile-to-sterile neutrino magnetic moment as a function of the relevant parameters of the UV model. The coloured areas correspond to the probed regions in Fig.~\ref{fig:results_B1}: $\delta=0.005$ in yellow, $\delta = 0.01$ in orange, $\delta= 0.05$ in red, $\delta=0.1$ in purple. The grey shaded area is excluded by direct collider searches. \textbf{Bottom panels}: Active-to-sterile neutrino magnetic moment as a function of the relevant parameters of the UV model. The pink ($m_{N_2}=80$~GeV) and blue ($m_{N_2}=700$~GeV) areas cover the probed dipole values in Fig.~\ref{fig:results_B2} for $C_{\s NNB}^{(5)}=3 \times 10^{-6}$~GeV$^{-1}$. The parameter space excluded by collider searches and cLFV processes is shown in grey for two flavour scenarios. The legend with the different constraints is displayed in the top right panel. }
\label{fig:link_model}
\end{figure}

Finally, we can map the values of the dipole couplings to which experiments are sensitive onto the parameters of the UV-complete model, namely the masses of the singly-charged scalar and vector-like lepton ($m_\phi, m_E$) and the Yukawa-type couplings in Eq.~\eqref{eq:model} ($f, f', Y_E, h, h'$). The sterile-to-sterile and active-to-sterile dipoles are related to these parameters via
\begin{align}
d_{\s{NN\gamma}}^{ij} = \frac{e}{128 \pi^2 m_E} \big(h_i' h_j^\ast - h_i^\ast h_j'\big)\, , \quad d_{\s{\nu N \gamma}}^{\alpha i} = \frac{e v}{32\sqrt{2}\pi^2m_E^2} f_{\alpha \beta} Y_E^{\beta \ast} h_i' \, ,
\end{align}
where we have assumed that $m_E=m_\phi$, hence replacing $f(r)\big|_{r\to 1} = \frac{1}{2}$. Since there are products of two or even three couplings entering these equations, we find it convenient to plot the generated magnetic moments as functions of the specific coupling combinations and the vector-like lepton mass. In the top panel of Fig.~\ref{fig:link_model}, we show fixed values of $d_{\s{NN\gamma}}$ ($d_{\s{NN\gamma}}^{12}$) ranging from $10^{-5}$ to $10^{-9}$~GeV$^{-1}$ in the corresponding plane.  Additionally, the coloured areas correspond to the probed regions in Fig.~\ref{fig:results_B1}, marginalising the dependence on $m_{N_2}$. Direct collider searches exclude the low mass region of $m_E$ (see discussion in Section~\ref{sec:constraints}).
The bottom panels in Fig.~\ref{fig:link_model} show active-to-sterile magnetic moments ranging from $10^{-7}$ to $10^{-13}$~GeV$^{-1}$. Here, the pink and blue areas cover the probed regions illustrated in Fig.~\ref{fig:results_B2} for the 30 event contours assuming $C_{\s NNB}^{(5)}= 10^{-5}$~GeV$^{-1}$ (which correspond to 3 event contours assuming $C_{\s NNB}^{(5)}=3\times 10^{-6}$~GeV$^{-1}$), again marginalising the dependence on $m_{N_1}$. In addition to the collider constraints, we show bounds from cLFV processes (see discussion in Section~\ref{sec:constraints}) for two different flavour scenarios. The left plot assumes $\mu$-$\tau$ couplings for the vector-like lepton 
($Y_E^{\mu}=Y_E^{\tau} \gg Y_E^e\approx 0$ and $f_{\mu\tau} \neq 0$), while the right plot assumes couplings only to  $\tau$ ($Y_E^{\tau}\gg Y_E^{e,\mu}\approx 0$ and  $f_{e\tau} = f_{\mu\tau} \neq 0 $). As we have fixed $C_{\s{NNB}}^{(5)}$, there is an upper limit on $m_E$ above which the couplings of the sterile-to-sterile magnetic moment enter the non-perturbative regime. Masses above the grey dashed line lead to $h,h' >\sqrt{4\pi}$.

\begin{figure}[t]
\centering
\includegraphics[width=0.49\textwidth]{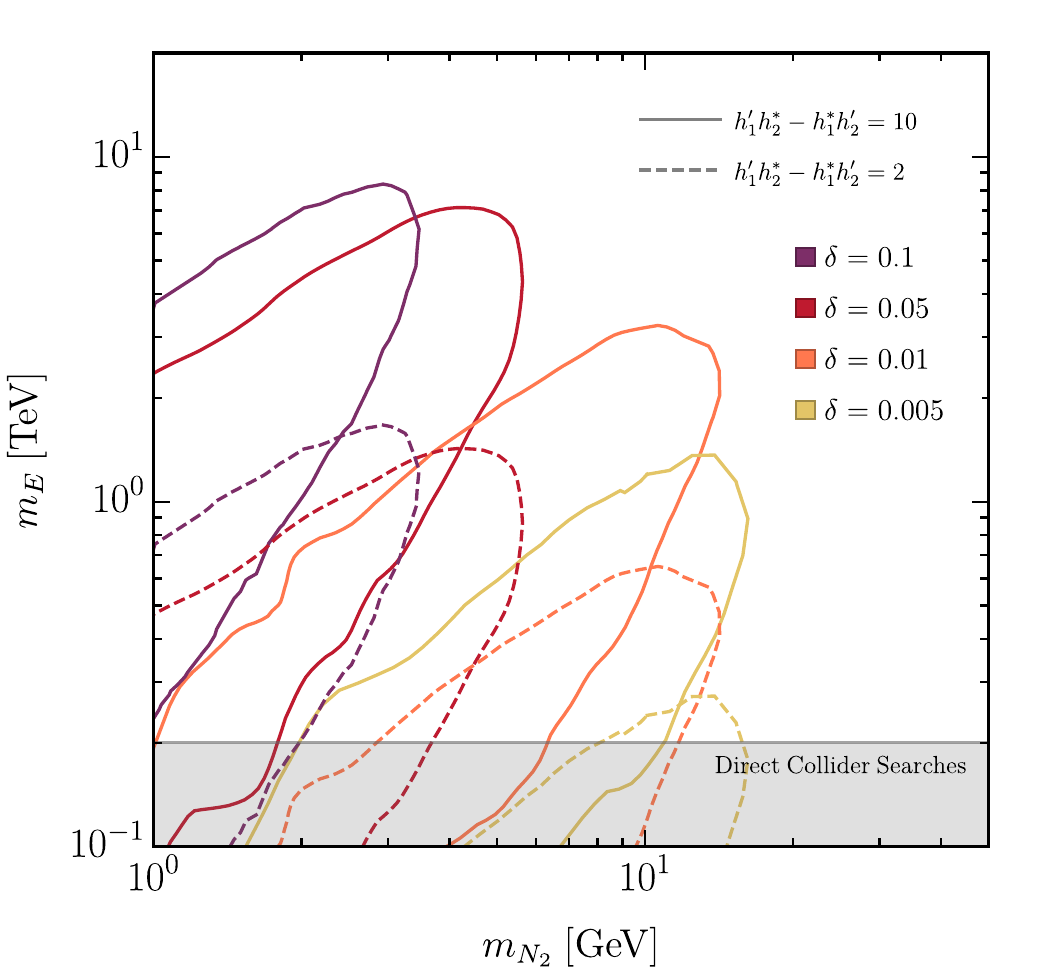}

\includegraphics[width=0.49\textwidth]{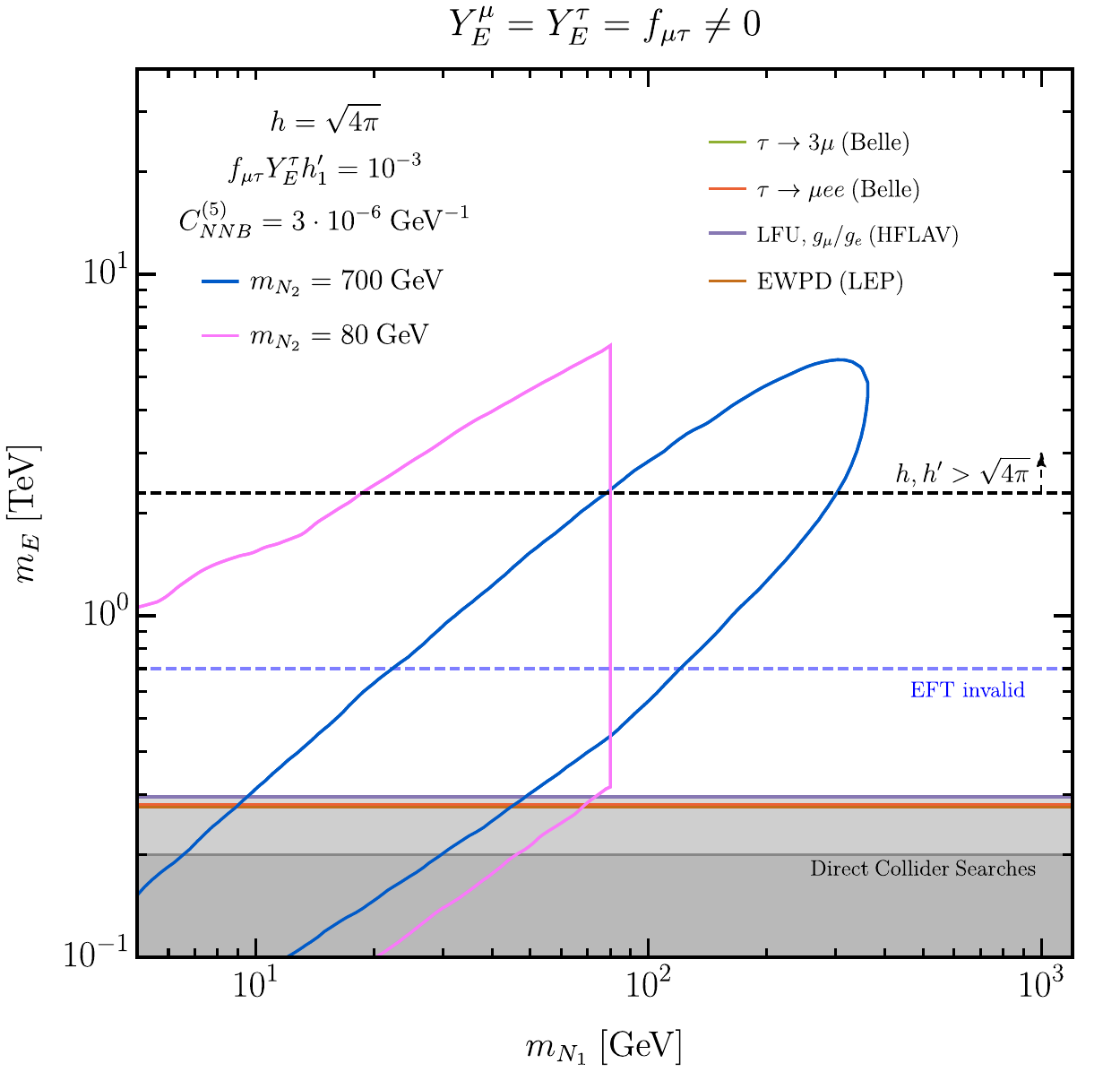} 
\includegraphics[width=0.49\textwidth]{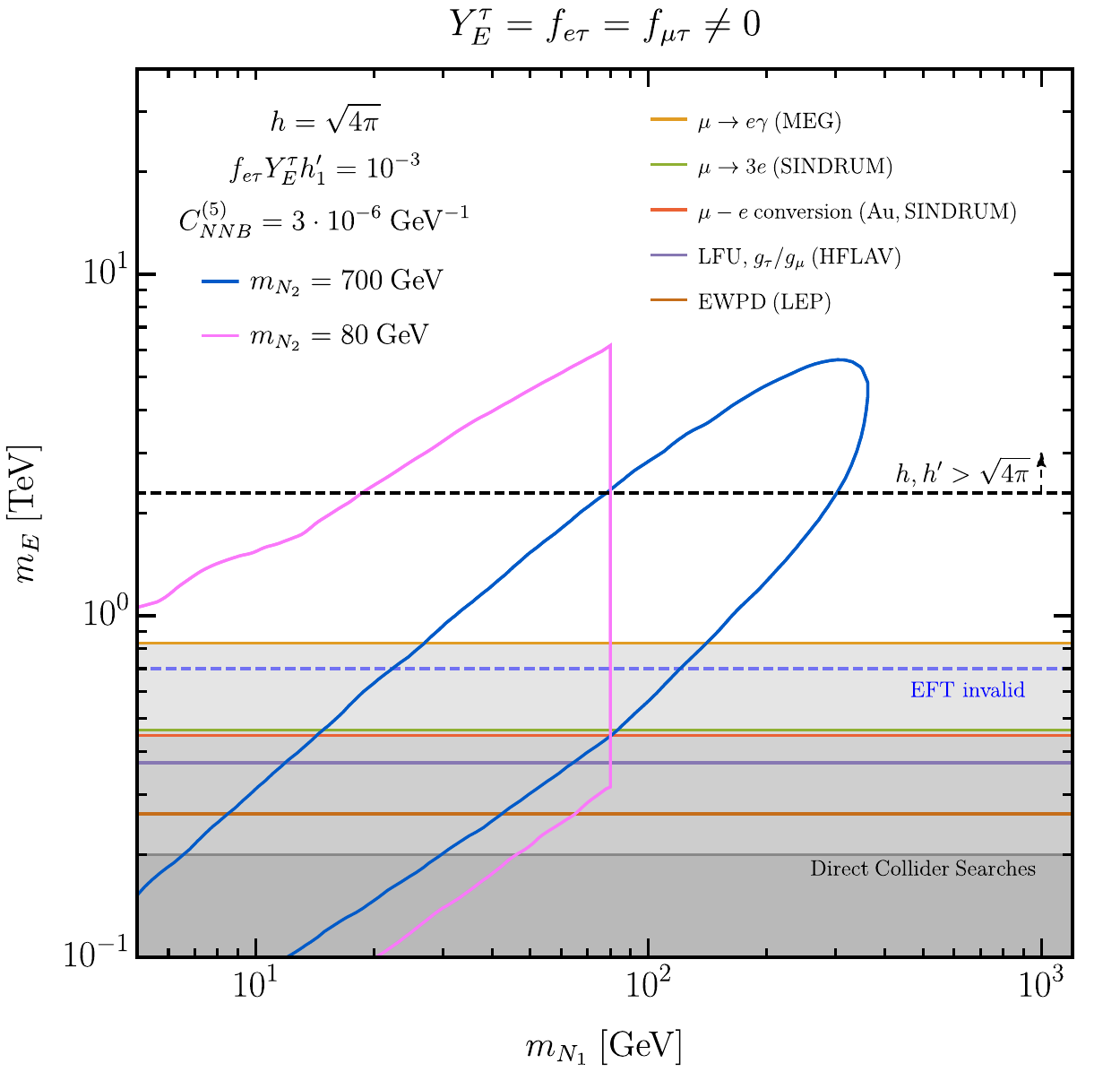} 
\caption{\textbf{Top panel}: Sensitivity projections from Fig.~\ref{fig:results_B1} in the plane $m_E$ vs.~$m_{N_2}$ for fixed coupling values. Direct collider searches exclude the lower grey band. \textbf{Bottom panels}: Sensitivity projections from Fig.~\ref{fig:results_B2} for $C_{\s NNB}^{(5)}=3\times10^{-6}$~GeV$^{-1}$ in the plane $m_E$ vs.~$m_{N_1}$. The left (right) plot assumes $\mu$--$\tau$ (only $\tau$) couplings of the vector-like lepton. Above the grey dashed line we enter the non-perturbative regime of $h,h'>\sqrt{4\pi}$, given the value of $C_{\s{NNB}}^{(5)}$. The lower grey regions are excluded by different cLFV processes and EWPD, which depend on the flavour scenario. }
\label{fig:mE_mN}
\end{figure}

In Fig.~\ref{fig:mE_mN}, we display in three different panels the sensitivity projections from Figs.~\ref{fig:results_B1} and \ref{fig:results_B2}. These are shown in the planes $m_E$ vs. $m_{N_2}$ for B1 (top panel) and $m_E$ vs. $m_{N_1}$ for B2 (bottom panels). We overlay current constraints from direct collider searches and cLFV processes as before. In the bottom panels, we also show the non-perturbative regime $h,h'>\sqrt{4\pi}$.
Additionally, the region corresponding to $m_E < m_{N_2} = 700$~GeV compromises the EFT validity. As can be seen from the top plot of Fig.~\ref{fig:mE_mN}, for scenario B1 (where the decay of sterile neutrinos is dominated by the sterile-to-sterile transition magnetic moment) LLP searches can be sensitive to vector-like lepton masses as high as several TeV (depending on the assumed value for the couplings) for sterile neutrino masses less than 10 GeV. However, the cLFV processes do not place additional constraints on this scenario because the couplings entering the cLFV rates are independent of the sterile-to-sterile transition magnetic moment.

In contrast, for scenario B2 (where the decay of sterile neutrinos is dominated by the active-to-sterile transition magnetic moment), the complementarity between LLP searches at the LHC and the constraints from cLFV becomes evident, as can be seen from the bottom two plots of Figs.~\ref{fig:link_model} and \ref{fig:mE_mN}. In this scenario, for a lepton flavour universal coupling of the vector-like lepton ($Y_E^e=Y_E^\mu=Y_E^\tau$), the stringent constraints from the current measurements of cLFV processes involving the first two generations of the charged leptons (combined with the perturbativity of Yukawa couplings) exclude the whole model parameter space. On the other hand, for the lepton flavour non-universal case, we have viable parameter space to be explored by LLP searches with non-pointing photons. If the vector-like lepton couples mainly to $\mu$  and $\tau$ with similar strength (with negligible coupling to the first-generation charged lepton, $ Y_E^{\mu}=Y_E^{\tau} \approx f_{\mu\tau} \gg Y_E^e\approx 0$) then the viable parameter space of the model in vector-like lepton mass spans between ($0.30-2.3$) TeV, with the lower limit coming from the current constraint from the $\tau\rightarrow \mu\gamma$ and the current measurement of LFY from tau decays. The upper limit corresponds to the perturbativity of the Yukawa couplings. If the vector-like lepton couples only to $\tau$ (with negligible coupling to the first two generations of charged lepton, $ Y_E^{\tau}\approx f_{e\tau} = f_{\mu\tau} \gg Y_E^{e,\mu}\approx 0$) then the viable parameter space of the model in vector-like lepton mass spans between ($0.8-2.3$) TeV, with the lower limit coming from the current constraints from the cLFV process $\mu\rightarrow e\gamma$ and the upper limit from the perturbativity of the Yukawa couplings.

\section{Conclusions}
\label{sec:conclusions}

In this work, we have examined long-lived particle (LLP) searches at the LHC using non-pointing photons to investigate sterile-to-sterile and active-to-sterile transition magnetic dipole moments. We considered two Majorana sterile neutrinos with masses ranging from a few GeV to several hundred GeV with sizeable sterile-to-sterile and active-to-sterile transition magnetic moments, in addition to the usual active-sterile neutrino mixing. We first discussed the operators relevant to neutrino magnetic moments within an EFT framework; specifically, $N_R$SMEFT and its low-energy counterpart, $N_R$LEFT. Subsequently, we considered as an example a simplified UV-complete model, illustrating the emergence of substantial transition magnetic moments between the heavy sterile neutrinos at the loop level. We discussed relevant constraints on this model from direct collider searches and charged lepton flavour violating processes.

For the phenomenological analysis of sterile neutrino magnetic moments, we began by taking the EFT approach. We considered the impact of two independent $N_R$SMEFT operators, $\mathcal{O}_{\s{NNB}}^{(5)}$ and $\mathcal{O}_{\s{NB}}^{(6)}$, together with the active-sterile neutrino mixing. We identified nine possible scenarios based on the dominant production and decay modes at the LHC. Of these nine possibilities, we explored the three physically realisable cases in detail. In these, the sterile-to-sterile magnetic moment dominates the production cross-section of sterile neutrinos at the LHC. Since there are no charged tracks from the primary vertex for the scenarios of interest, we have put forward a search strategy employing the transverse impact parameter of non-pointing photons, which does not rely on the location of the primary vertex.

We presented detailed sensitivities for the three physically realisable scenarios at the high-luminosity LHC. Our numerical simulations indicate that for sterile neutrinos decaying primarily to photons, either via sterile-to-sterile (scenario B1) or active-to-sterile magnetic dipole moments (scenario B2), searches for LLPs with non-pointing photons can probe unexplored regions of the parameter space for sterile neutrino masses ranging from a few GeV to several hundred GeV. We find that the high-luminosity LHC will improve constraints on the sterile-to-sterile neutrino transition magnetic moment with respect to LEP searches by more than an order of magnitude for the sub-ten-GeV sterile neutrino mass regime. On the other hand, for scenario B2, where considerably more energetic photons are produced, the active-to-sterile neutrino magnetic moment can be probed to much lower values compared to existing constraints. In this case, an unprecedented reach for sterile neutrino masses up to hundreds of GeV is expected. For the scenario where the lighter sterile neutrino decays dominantly through mixing, our results reveal that LLP searches using displaced vertices could probe active-sterile neutrino mixing values much lower than those in the minimal scenario (where mixing controls both the production cross-sections and sterile neutrino lifetimes).

Applying the results from the model-independent EFT approach to our realistic simplified model example, we found that the synergy between LLP searches using non-pointing photons at the LHC and the constraints from cLFV processes can potentially probe and distinguish different flavour structures of new physics couplings. For the scenario where the decay of sterile neutrinos is dominated by the sterile-to-sterile transition magnetic moment, LLP searches will be sensitive to vector-like lepton masses as high as several TeV. However, such a scenario remains largely unconstrained from the cLFV processes owing to the freedom of choice for some of the couplings entering the cLFV rates, which are independent of the sterile-to-sterile transition magnetic moment. In contrast, for the scenario where the decay of sterile neutrinos is dominated by the active-to-sterile transition magnetic moment, the complementarity of LLP searches at the LHC and the constraints from cLFV can play a pivotal role in understanding the flavour structure of the model parameter space. For flavour universal couplings of the vector-like lepton, the stringent constraints from the current measurements of cLFV processes involving the first two generations of charged leptons already exclude the whole viable model parameter space. On the other hand, if the vector-like lepton couples mainly to $\mu$ and (or) $\tau$, future cLFV measurements will also provide excellent complementarity with LLP searches at the LHC using non-pointing photons.


\paragraph {\textbf{Acknowledgements}}
R.~B. is supported by the grants ACIF/2021/052 and CIBEFP/2022/62 (Generalitat Valenciana). P.~D.~B. is supported by the Slovenian Research Agency under the research core funding No. P1-0035 and in part by the research grants N1-0253 and J1-4389. P.~D.~B. also acknowledges support from the European Union's Horizon 2020 research and innovation programme under the Marie Sk\l{}odowska-Curie grant agreement No 860881-HIDDeN. F.~F.~D. acknowledges support from the UK Science and Technology Facilities Council (STFC) via the Consolidated Grants ST/P00072X/1 and ST/T000880/1. C.~H. is supported by the Generalitat Valenciana under Plan Gen-T via CDEIGENT grant No. CIDEIG/2022/16. R.~B. also acknowledges partial support from CDEIGENT grant No. CIDEIG/2022/16. R.~B., C.~H., and M.~H. acknowledge partial support from the Spanish grants PID2020- 113775GBI00 (AEI/10.13039/501100011033), and Prometeo
CIPROM/2021/054 (Generalitat Valenciana).  

\bibliographystyle{JHEP}
\bibliography{bibliography}
\end{document}